\documentclass[12pt]{article}
\usepackage{graphicx}

\usepackage{epsf}
\def\beq{\begin{equation}}
\def\eeq{\end{equation}}
\def\bea{\begin{eqnarray}}
\def\eea{\end{eqnarray}}
\def\ve{\vert}
\def\vel{\left|}
\def\ver{\right|}
\def\nnb{\nonumber}
\def\ga{\left(}
\def\dr{\right)}
\def\aga{\left\{}
\def\adr{\right\}}
\def\rar{\rightarrow}
\def\nnb{\nonumber}
\def\la{\langle}
\def\ra{\rangle}
\def\ba{\begin{array}}
\def\ea{\end{array}}
\def\bea{\begin{eqnarray}}
\def\eea{\end{eqnarray}}
\def\tep{$b \rar s \ell^+ \ell^-$}

\def\ds{\displaystyle}
\def\ve{\vert}
\def\vel{\left|}
\def\ver{\right|}
\def\nnb{\nonumber}
\def\ga{\left(}
\def\dr{\right)}
\def\aga{\left\{}
\def\adr{\right\}}
\def\rar{\rightarrow}
\def\nnb{\nonumber}
\def\la{\langle}
\def\ra{\rangle}
\def\lla{\left<}
\def\rra{\right>}
\def\simlt{\stackrel{<}{{}_\sim}}
\def\simgt{\stackrel{>}{{}_\sim}}
\begin{document}
\def\beq{\begin{equation}}
\def\eeq{\end{equation}}
\def\bea{\begin{eqnarray}}
\def\eea{\end{eqnarray}}
\def\ve{\vert}
\def\vel{\left|}
\def\ver{\right|}
\def\nnb{\nonumber}
\def\ga{\left(}
\def\dr{\right)}
\def\aga{\left\{}
\def\adr{\right\}}
\def\rar{\rightarrow}
\def\nnb{\nonumber}
\def\la{\langle}
\def\ra{\rangle}
\def\lla{\left<}
\def\rra{\right>}
\def\ba{\begin{array}}
\def\ea{\end{array}}
\def\BcDll{$B_c \rar D_{s}^* \ell^+ \ell^-$}
\def\tepm{$B \rar K \mu^+ \mu^-$}
\def\tept{$B \rar K \tau^+ \tau^-$}
\def\ds{\displaystyle}



\newskip\humongous \humongous=0pt plus 1000pt minus 1000pt
\def\caja{\mathsurround=0pt}
\def\eqalign#1{\,\vcenter{\openup1\jot
\caja   \ialign{\strut \hfil$\displaystyle{##}$&$
\displaystyle{{}##}$\hfil\crcr#1\crcr}}\,}


\def\simlt{\stackrel{<}{{}_\sim}}
\def\simgt{\stackrel{>}{{}_\sim}}



\def\bos{\lower 0.5cm\hbox{{\vrule width 0pt height 1.2cm}}}
\def\boss{\lower 0.35cm\hbox{{\vrule width 0pt height 1.cm}}}
\def\aaa{\lower 0.cm\hbox{{\vrule width 0pt height .7cm}}}
\def\dol{\lower 0.4cm\hbox{{\vrule width 0pt height .5cm}}}


\title{ {\Large {\bf
General analysis of the rare \BcDll decay beyond the standard
model } } }

\author
{{\small U. O. Yilmaz $^1$ \, and \, G. Turan $^2$ \thanks {e-mail: gsevgur@metu.edu.tr} }\\
 \\{\small $^1$ Physics Department, Mersin University 33343 Ciftlikkoy Mersin, Turkey} \\
 {\small $^2$ Physics Department, Middle East Technical University 06531 Ankara, Turkey}
 }
\date { }

\begin{titlepage}
\maketitle
\thispagestyle{empty}

\begin{abstract}
The general analysis of the  rare \BcDll decay is presented  by
using the most general, model independent effective Hamiltonian. The
dependencies  of  the branching ratios, longitudinal, normal and
transversal polarization asymmetries for $\ell^-$  and the combined
asymmetries for $\ell^-$ and $\ell^+$ on the new Wilson coefficients
are investigated.  Our analysis shows that the lepton polarization
asymmetries are very sensitive to the scalar and tensor type
interactions, which will be very useful in looking for new physics
beyond the standard model.
\end{abstract}

\end{titlepage}

\section{Introduction}
Rare $B$ meson decays, induced by flavor--changing neutral current
(FCNC) $b \rar s,d$ transitions, are excellent places to search
for new physics because they appear at the same order as the
standard model (SM). In rare $B$ meson decays, effects of the new
physics may appear in two different manners, either through the
new contributions to the Wilson coefficients existing in the SM or
through the new structures in the effective Hamiltonian which are
absent in the SM.

There have been many investigations of the new physics through the
study of rare radiative, leptonic and semileptonic decays of
$B_{u,d,s}$ mesons induced by FCNC transitions of $b \rar s, d$
\cite{rev} since the CLEO observation of $b \rar s \, \gamma$
\cite{CLEO}. The studies will be even more complete if similar
decays for $B_c$ are also included.

The study of the $B_c$ meson is by itself quite interesting too,
due to its outstanding features \cite{Colangelo}-\cite{Geng} . It
is the lowest bound state of two heavy quarks ($b$ and $c$) with
explicit flavor that can be compared with the charmonium
($c\bar{c}$- bound state) and bottomium ($b\bar{b}$- bound state)
which have implicit flavor. The implicit-flavor states decay
strongly and electromagnetically whereas the $B_c$ meson decays
weakly. The major difference between the weak decay properties of
$B_c$ and $B_{u,d,s}$ is that those of the latter ones are
described very well in the framework of the heavy quark limit,
which gives some relations between the form factors of the
physical process. In case of $B_c$ meson, the heavy flavor and
spin symmetries must be reconsidered because both $b$ and $c$ are
heavy.

From the experimental side, the running B factories in KEK and
SLAC continue to collect data samples and encourage the study of
rare B meson decays. It is believed that in future experiments at
hadronic colliders, such as the BTeV and LHC-B   most of rare
$B_c$ decays should be accessible.

One of the efficient ways in establishing new physics beyond the
SM is the measurement of the lepton polarization
\cite{R15}--\cite{R24}. In this work we present a  study of the
branching ratio and lepton polarizations in the exclusive \BcDll
 $~(\ell = \mu,~\tau)$ decay for a general form of the
effective Hamiltonian including all possible form of interactions
in a model independent way  without forcing concrete values for
the Wilson coefficients corresponding to any specific model.

It is well known that the theoretical study of the inclusive decays is
rather easy but their experimental investigation is difficult. However for
the exclusive decays the situation is contrary to the inclusive case, i.e.,
their experimental detection is very easy but theoretical investigation
has its own drawbacks. This is due to the fact that for description of the exclusive
decay form factors, i.e., the matrix elements of the effective Hamiltonian
between initial and final meson states, are needed. This problem is related
to the nonperturbative sector of the QCD and and it can only be solved in
framework of the nonperturbative approaches.

These matrix elements have been studied in framework of different
approaches, such as light front, constituent quark models \cite{Geng},  and
a relativistic quark model proposed in ref.  \cite{Faessler}. In this work we will
use the weak decay form factors calculated in ref.  \cite{Faessler}.

The paper is organized as follows. In section 2, we first give the
effective Hamiltonian for the quark level process \tep and the
definitions of the form factors, and then introduce the
corresponding matrix element. In section 3, we present   the model
independent expressions for the longitudinal, transversal and
normal polarizations of leptons. We also give the
lepton-antilepton combined asymmetries. Section 4 is devoted to
the numerical analysis and discussion of our results.

\section{Effective Hamiltonian}

In the standard effective Hamiltonian approach,  the  \BcDll decay is described at the quark level
by the $b \rar  s \ell^+ \ell^- $ process, which can be written in terms of twelve model
independent four-Fermi interactions as follows  \cite{R19},
\bea
\label{effH}
{\cal H}_{eff} &=& \frac{G\alpha}{\sqrt{2} \pi}
 V_{ts}V_{tb}^\ast
\Bigg\{ C_{SL} \, \bar s i \sigma_{\mu\nu} \frac{q^\nu}{q^2}\, L \,b
\, \bar \ell \gamma^\mu \ell + C_{BR}\, \bar s i \sigma_{\mu\nu}
\frac{q^\nu}{q^2} \,R\, b \, \bar \ell \gamma^\mu \ell \nnb \\
&+&C_{LL}^{tot}\, \bar s_L \gamma_\mu b_L \,\bar \ell_L \gamma^\mu \ell_L +
C_{LR}^{tot} \,\bar s_L \gamma_\mu b_L \, \bar \ell_R \gamma^\mu \ell_R +
C_{RL} \,\bar s_R \gamma_\mu b_R \,\bar \ell_L \gamma^\mu \ell_L \nnb \\
&+&C_{RR} \,\bar s_R \gamma_\mu b_R \, \bar \ell_R \gamma^\mu \ell_R +
C_{LRLR} \, \bar s_L b_R \,\bar \ell_L \ell_R +
C_{RLLR} \,\bar s_R b_L \,\bar \ell_L \ell_R \\
&+&C_{LRRL} \,\bar s_L b_R \,\bar \ell_R \ell_L +
C_{RLRL} \,\bar s_R b_L \,\bar \ell_R \ell_L+
C_T\, \bar s \sigma_{\mu\nu} b \,\bar \ell \sigma^{\mu\nu}\ell \nnb \\
&+&i C_{TE}\,\epsilon^{\mu\nu\alpha\beta} \bar s \sigma_{\mu\nu} b \,
\bar \ell \sigma_{\alpha\beta} \ell  \Bigg\}~, \nnb
\eea
where the chiral projection operators $L$ and $R$ in (\ref{effH}) are
defined as
\bea
L = \frac{1-\gamma_5}{2} ~,~~~~~~ R = \frac{1+\gamma_5}{2}\nnb~,
\eea
and $C_X$ are the coefficients of the four--Fermi interactions. The  coefficients, $C_{SL}$ and $C_{BR}$,
are the nonlocal Fermi interactions which correspond to $-2 m_s C_7^{eff}$ and $-2 m_b C_7^{eff}$
in the SM, respectively. The following four terms in Eq. (\ref{effH}) are the vector type interactions with
coefficients $C_{LL}$, $C_{LR}$, $C_{RL}$ and $C_{RR}$. Two of these
vector interactions containing $C_{LL}^{tot}$ and $C_{LR}^{tot}$  already exist in the SM
in combinations of the form $(C_9^{eff}-C_{10})$ and $(C_9^{eff}+C_{10})$.
Therefore we write
\bea
C_{LL}^{tot} &=& C_9^{eff} - C_{10} + C_{LL}~, \nnb \\
C_{LR}^{tot} &=& C_9^{eff} + C_{10} + C_{LR}~, \nnb
\eea
so that $C_{LL}^{tot}$ and $C_{LR}^{tot}$ describe the
sum of the contributions from SM and the new physics. The terms with
coefficients $C_{LRLR}$, $C_{RLLR}$, $C_{LRRL}$ and $C_{RLRL}$ describe
the scalar type interactions. The remaining two terms with the
coefficients $C_T$ and $C_{TE}$ describe the tensor type interactions.

After giving the general form of four--Fermi interaction for the $b
\rar s \ell^+ \ell^-$ transition, we now need to estimate  the
matrix element for the \BcDll decay. These can be expressed in term
of invariant form factors as follows:
\bea
\lefteqn{ \label{ilk}
\lla D_{s}^\ast(p_{D^\ast},\varepsilon) \vel \bar s \gamma_\mu
(1 \pm \gamma_5) b \ver B_c(p_{B_c}) \rra =} \nnb \\
&&- \epsilon_{\mu\nu\lambda\sigma} \varepsilon^{\ast\nu} p_{D^\ast}^\lambda q^\sigma
\frac{2 V(q^2)}{m_{B_c}+m_{D^\ast}} \pm i \varepsilon_\mu^\ast (m_{B_c}-m_{D^\ast})
A_0(q^2) \\
&&\mp i (p_{B_c} + p_{D^\ast})_\mu (\varepsilon^\ast q)
\frac{A_+(q^2)}{m_{B_c}+m_{D^\ast}}
\mp i q_\mu (\varepsilon^\ast q)\frac{A_-(q^2)}{m_{B_c}+m_{D^\ast}} ~,  \nnb \\  \nnb \\
\lefteqn{ \label{iki} \lla D_{s}^\ast(p_{D^\ast},\varepsilon) \vel
\bar s i \sigma_{\mu\nu} q^\nu
(1 \pm \gamma_5) b \ver B(p_{B_c}) \rra =} \nnb \\
&&2 \epsilon_{\mu\nu\lambda\sigma} \varepsilon^{\ast\nu} p_{D^\ast}^\lambda q^\sigma
g(q^2) \pm  i \left[ \varepsilon_\mu^\ast (m_{B_c}^2-m_{D^\ast}^2) -
(p_{B_c} + p_{D^\ast})_\mu (\varepsilon^\ast q) \right] a_0(q^2) \\
&&\pm  i (\varepsilon^\ast q) \left[ q_\mu -
(p_{B_c} + p_{D^\ast})_\mu \frac{q^2}{m_{B_c}^2-m_{D^\ast}^2} \right]
\frac{ (m_{B_c}^2-m_{D^\ast}^2)}{q^2}(a_+(q^2) - a_0(q^2))~, \nnb \\  \nnb \\
\lefteqn{ \label{ucc}
\lla D_{s}^\ast(p_{D^\ast},\varepsilon) \vel
\bar s \sigma_{\mu\nu}
 b \ver B(p_{B_c}) \rra =} \nnb \\
&&i \epsilon_{\mu\nu\lambda\sigma}  \Bigg[ -  g(q^2)
{\varepsilon^\ast}^\lambda (p_{B_c} + p_{D^\ast})^\sigma +
\frac{1}{q^2} (m_{B_c}^2-m_{D^\ast}^2) {\varepsilon^\ast}^\lambda
q^\sigma (g(q^2)-a_0(q^2))\\
&&- \frac{2}{q^2} \Bigg(g(q^2) - a_+(q^2) \Bigg) (\varepsilon^\ast q) p_{D^\ast}^\lambda q^\sigma \Bigg]~. \nnb
\eea
\bea \lla D_{s}^\ast(p_{D^\ast},\varepsilon) \vel \bar s (1 \pm
\gamma_5) b \ver B(p_{B_c}) \rra & =
& \frac{1}{m_b} \Big[ \mp  (\varepsilon^\ast q)(m_{B_c} - m_{D_{s}^\ast}) \nnb \\
&+& A_0(q^2) - A_{+}(q^2) - \frac{q^2}{m_{B_c}^2 -
    m_{D^\ast}^2} A_{-}(q^2)  \Big]~.\label{uc} \eea
where $q = p_{B_c}-p_{D_{s}^\ast}$ is the momentum transfer and
$\varepsilon$ is the polarization vector of $D_{s}^\ast$ meson. The
matrix element $\lla D_{s}^\ast \vel \bar s (1 \pm \gamma_5 ) b \ver
B \rra$ is calculated  by contracting both sides of Eq. (\ref{ilk})
with $q^\mu$ and using equation of motion.

By using  Eqs. (\ref{effH})--(\ref{uc}), we can now write the matrix
element of the \BcDll decay as \bea \lefteqn{ \label{had}
    {\cal M}(B_c\rightarrow D_{s}^\ast \ell^{+}\ell^{-}) =
    \frac{G \alpha}{4 \sqrt{2} \pi} V_{tb} V_{ts}^\ast }\nnb \\
&&\times \Bigg\{
    \bar \ell \gamma^\mu(1-\gamma_5) \ell \, \Big[
    -2 A_1 \epsilon_{\mu\nu\lambda\sigma} \varepsilon^{\ast\nu}
    p_{D_{s}^\ast}^\lambda q^\sigma
    -i B_1 \varepsilon_\mu^\ast
    + i B_2 (\varepsilon^\ast q) (p_{B_c}+p_{D_{s}^\ast})_\mu
    + i B_3 (\varepsilon^\ast q) q_\mu  \Big] \nnb \\
&&+ \bar \ell \gamma^\mu(1+\gamma_5) \ell \, \Big[
    -2 C_1 \epsilon_{\mu\nu\lambda\sigma} \varepsilon^{\ast\nu}
    p_{D_{s}^\ast}^\lambda q^\sigma
    -i D_1 \varepsilon_\mu^\ast
    + i D_2 (\varepsilon^\ast q) (p_{B_c}+p_{D_{s}^\ast})_\mu
    + i D_3 (\varepsilon^\ast q) q_\mu  \Big] \nnb \\
&&+\bar \ell (1-\gamma_5) \ell \Big[ i B_4 (\varepsilon^\ast
    q)\Big]
    + \bar \ell (1+\gamma_5) \ell \Big[ i B_5 (\varepsilon^\ast
    q)\Big]  \nnb \\
&&+4 \bar \ell \sigma^{\mu\nu}  \ell \Big( i C_T \epsilon_{\mu\nu\lambda\sigma}
    \Big) \Big[ -2 g {\varepsilon^\ast}^\lambda (p_{B_c}+p_{D_{s}^\ast})^\sigma +
    B_6 {\varepsilon^\ast}^\lambda q^\sigma -
    B_7 (\varepsilon^\ast q) {p_{D_{s}^\ast}}^\lambda q^\sigma \Big] \nnb \\
&&+16 C_{TE} \bar \ell \sigma_{\mu\nu}  \ell \Big[ -2 g
    {\varepsilon^\ast}^\mu (p_{B_c}+p_{D_{s}^\ast})^\nu  +B_6 {\varepsilon^\ast}^\mu q^\nu -
    B_7 (\varepsilon^\ast q) {p_{D_{s}^\ast}}^\mu q^\nu
    \Bigg\}~,
\eea where \bea \label{as} A_1 &=& (C_{LL}^{tot} + C_{RL})
\frac{V}{m_{B_c}+m_{D_{s}^\ast}} -
    (C_{BR}+C_{SL}) \frac{g}{q^2} ~, \nnb \\
B_1 &=& (C_{LL}^{tot} - C_{RL}) (m_{B_c}-m_{D_{s}^\ast}) A_0 -
    (C_{BR}-C_{SL}) (m_{B_c}^2-m_{D_{s}^\ast}^2)
    \frac{a_0}{q^2} ~, \nnb \\
B_2 &=& \frac{C_{LL}^{tot} - C_{RL}}{m_{B_c}+m_{D_{s}^\ast}} A_+ -
    (C_{BR}-C_{SL}) \frac{a_+}{q^2}  ~,
    \nnb \\
B_3 &=&  (C_{LL}^{tot} - C_{RL})
\frac{A_-}{m_{B_c}+m_{D_{s}^\ast}}+
    2 (C_{BR}-C_{SL}) (a_{+} - a_{0})\frac{ m_{B_c}^2 - m_{D_{s}^\ast}^2 }{q^4} ~, \nnb
    \\
B_4 &=& -  ( C_{LRRL} - C_{RLRL}) \frac{m_{B_c}-m_{D_{s}^\ast}}{m_b}
    \Big(A_0-A_- -\frac{q^2}{m_{B_c}^2-m_{D_{s}^\ast}^2}\Big) ~,\nnb \\
B_5 &=& -  ( C_{LRLR} - C_{RLLR}) \frac{m_{B_c}-m_{D_{s}^\ast}}{m_b}
    \Big(A_0-A_- -\frac{q^2}{m_{B_c}^2-m_{D_{s}^\ast}^2}\Big)  ~,\nnb \\
B_6 &=&  (m_{B_c}^2-m_{D_{s}^\ast}^2) \frac{g-a_0}{q^2} ~,\nnb \\
B_7 &=& \frac{2}{q^2} \left( g-a_0 -
    a_+\right)~,\nnb \\
C_1 &=& A_1 ( C_{LL}^{tot} \rar C_{LR}^{tot}~,~~C_{RL} \rar
    C_{RR})~,\nnb \\
    D_1 &=& B_1 ( C_{LL}^{tot} \rar C_{LR}^{tot}~,~~C_{RL} \rar
    C_{RR})~,\nnb \\
D_2 &=& B_2 ( C_{LL}^{tot} \rar C_{LR}^{tot}~,~~C_{RL} \rar
    C_{RR})~,\nnb \\
D_3 &=& B_3 ( C_{LL}^{tot} \rar C_{LR}^{tot}~,~~C_{RL} \rar
    C_{RR})~. \eea
\section{Lepton polarizations}
We now like to  calculate  the final lepton polarizations  for the
$B_c \rar D_{s}^\ast \ell^+ \ell^-$ decay. For this  we will use the
convention followed by the earlier works, such as
\cite{R18},\cite{R19} and  define the following orthogonal unit
vectors, $S_i^{-\mu}$ in the rest frame of $\ell^-$ and $S_i^{+\mu}$
in the rest frame of $\ell^+$, for the polarization of the leptons
along the longitudinal ($i=L$), transverse  ($i=T$) and normal
($i=N$) directions: \bea \label{pol} S_L^{-\mu} &\equiv&
(0,\vec{e}_L^{\,-}) =
\ga 0,\frac{\vec{p}_-}{\vel \vec{p}_- \ver} \dr~, \nnb \\
S_N^{-\mu} &\equiv& (0,\vec{e}_N^{\,-}) =
\ga 0,\frac{\vec{p} \times \vec{p}_-}
{\vel \vec{p} \times \vec{p}_- \ver} \dr~, \nnb \\
S_T^{-\mu} &\equiv& (0,\vec{e}_T^{\,-}) =
\ga 0, \vec{e}_N^{\,-} \times \vec{e}_L^{\,-} \dr~, \\
S_L^{+\mu} &\equiv& (0,\vec{e}_L^{\,+}) =
\ga 0,\frac{\vec{p}_+}{\vel \vec{p}_+ \ver} \dr~, \nnb \\
S_N^{+\mu} &\equiv& (0,\vec{e}_N^{\,+}) =
\ga 0,\frac{\vec{p} \times \vec{p}_+}
{\vel \vec{p} \times \vec{p}_+ \ver} \dr~, \nnb \\
S_T^{+\mu} &\equiv& (0,\vec{e}_T^{\,+}) = \ga 0, \vec{e}_N^{\,+}
\times \vec{e}_L^{\,+} \dr~, \nnb \eea where $\vec{p}_\pm$ and
$\vec{p}$ are the three momenta of $\ell^\pm$ and $D_{s}^\ast$ meson
in the center of mass (CM) frame of the $\ell^+ \ell^-$ system,
respectively. The longitudinal unit vectors $S_L^-$ and $S_L^+$ are
boosted to CM frame of $\ell^+ \ell^-$ by Lorentz transformation,
\bea \label{bs} S^{-\mu}_{L,\, CM} &=& \ga \frac{\vel \vec{p}_-
\ver}{m_\ell},
\frac{E_\ell \,\vec{p}_-}{m_\ell \vel \vec{p}_- \ver} \dr~, \nnb \\
S^{+\mu}_{L,\, CM} &=& \ga \frac{\vel \vec{p}_- \ver}{m_\ell},
- \frac{E_\ell \, \vec{p}_-}{m_\ell \vel \vec{p}_- \ver} \dr~,
\eea
while vectors of perpendicular directions are not changed by boost.

The differential decay rate of the $B_c \rar D_{s}^\ast \ell^+
\ell^-$ decay for any spin direction $\vec{n}^{\pm}$ of the
$\ell^{\pm}$ can be written in the following form \bea \label{ddr}
\frac{d\Gamma(\vec{n}^{\pm})}{ds} = \frac{1}{2} \ga
\frac{d\Gamma}{ds}\dr_0 \Bigg[ 1 + \Bigg( P_L^{\pm}
\vec{e}_L^{\,\pm} + P_N^{\pm} \vec{e}_N^{\,\pm} + P_T^{\pm}
\vec{e}_T^{\,\pm} \Bigg) \cdot \vec{n}^{\pm} \Bigg]~, \eea where
$\vec{n}^{\,\pm}$ is the unit vector in the $\ell^{\pm}$ rest frame
and $s=q^2/m_{B_c}^2$. Here, the superscripts $^+$ and $^-$
correspond to $\ell^+$ and $\ell^-$ cases, the subscript $_0$
corresponds to the unpolarized decay rate, whose explicit form is
given by \bea \label{unp} \lefteqn{ \ga \frac{d \Gamma}{ds}\dr_0 =
\frac{G^2 \alpha^2 m_{B_c}}{2^{14} \pi^5 }
\vel V_{tb} V_{ts}^\ast \ver^2 \sqrt{\lambda}\, v} \nnb \\
&\times& \Bigg\{ \frac{32}{3} m_{B_c}^4 \lambda \Big[(m_{B_c}^2 s -
    m_\ell^2) \ga \vel  A_1 \ver^2 + \vel  C_1 \ver^2 \dr + 6 m_\ell^2
    \, \mbox{\rm Re}
    (A_1 C_1^\ast)\Big] \nnb \\
&+& 96 m_\ell^2 \, \mbox{\rm Re} (B_1 D_1^\ast)-
    \frac{4}{r} m_{B_c}^2 m_\ell \lambda \,
    \mbox{\rm Re} [(B_1 - D_1) (B_4^\ast - B_5^\ast)] \nnb \\
&+&\frac{8}{r} m_{B_c}^2 m_\ell^2 \lambda \, \Big(
    \mbox{\rm Re} [B_1 (- B_3^\ast + D_2^\ast + D_3^\ast)] +
    \mbox{\rm Re} [D_1 (B_2^\ast + B_3^\ast - D_3^\ast)] -
    \mbox{\rm Re}(B_4 B_5^\ast) \Big] \Big)~~~~~~~~ \nnb \\
&+&\frac{4}{r} m_{B_c}^4 m_\ell (1-r) \lambda \,
    \Big(\mbox{\rm Re} [(B_2 - D_2) (B_4^\ast - B_5^\ast)]
    +2 m_\ell \, \mbox{\rm Re} [(B_2 - D_2) (B_3^\ast - D_3^\ast)]
    \Big) \nnb \\
&-& \frac{8}{r}m_{B_c}^4 m_\ell^2 \lambda (2+2 r-s)\, \mbox{\rm Re} (B_2 D_2^\ast)
    +\frac{4}{r} m_{B_c}^4 m_\ell s \lambda \,
    \mbox{\rm Re} [(B_3 - D_3) (B_4^\ast - B_5^\ast)] \nnb \\
&+&\frac{4}{r} m_{B_c}^4 m_\ell^2 s \lambda \,
    \vel B_3 - D_3\ver^2
    +\frac{2}{r} m_{B_c}^2 (m_{B_c}^2 s-2 m_\ell^2) \lambda \,
    \ga \vel B_4 \ver^2 + \vel B_5 \ver^2 \dr \nnb \\
&-&\frac{8}{3rs} m_{B_c}^2 \lambda \,
    \Big[m_\ell^2 (2-2 r+s)+m_{B_c}^2 s (1-r-s) \Big]
    \Big[\mbox{\rm Re}(B_1 B_2^\ast) + \mbox{\rm Re}(D_1 D_2^\ast)\Big] \nnb \\
&+&\frac{4}{3rs}\,
    \Big[2 m_\ell^2 (\lambda-6 rs)+m_{B_c}^2 s (\lambda+12 rs) \Big]
    \ga \vel B_1 \ver^2 + \vel D_1 \ver^2 \dr \nnb \\
&+&\frac{4}{3rs} m_{B_c}^4 \lambda\,
    \Big( m_{B_c}^2 s \lambda + m_\ell^2 [ 2 \lambda + 3 s (2+2 r - s) ] \Big)
    \ga \vel B_2 \ver^2 + \vel D_2 \ver^2 \dr \nnb \\
&+&\frac{32}{r} m_{B_c}^6 m_\ell \lambda^2 \,
    \mbox{\rm Re} [(B_2 + D_2)(B_7 C_{TE})^\ast]  \\
&-& \frac{32}{r} m_{B_c}^4 m_\ell \lambda (1-r-s) \Big(
    \mbox{\rm Re} [(B_1 + D_1)(B_7 C_{TE})^\ast] +
    2\, \mbox{\rm Re} [(B_2 + D_2)(B_6 C_{TE})^\ast] \Big) \nnb \\
&+&\frac{64}{r} (\lambda+12 rs) m_{B_c}^2 m_\ell \,
    \mbox{\rm Re} [(B_1 + D_1)(B_6 C_{TE})^\ast] \nnb \\
&+&\frac{256}{3rs} \vel g \ver^2 \vel C_T \ver^2 m_{B_c}^2
    \Big( 4 m_\ell^2 \, [ \lambda ( 8r -s) - 12 r s (2 +2r -s) ] \nnb \\
&+& m_{B_c}^2 s \, [\lambda (16r+s)+12 r s (2 +2r -s) ] \Big) \nnb \\
&+&\frac{1024}{3rs} \vel g \ver^2 \vel C_{TE} \ver^2 m_{B_c}^2
    \Big( 8 m_\ell^2 \, [ \lambda ( 4r+s) + 12 r s (2 +2r -s) ] \nnb \\
&+& m_{B_c}^2 s \, [\lambda (16r+s)+12 r s (2 +2r -s) ] \Big) \nnb \\
&-& \frac{128}{r} m_{B_c}^2 m_\ell \, [\lambda + 12 r (1-r) ]
    \,\mbox{\rm Re} [(B_1 + D_1)(g C_{TE})^\ast] \nnb \\
&+&\frac{128}{r} m_{B_c}^4 m_\ell \lambda (1+3r-s)
    \,\mbox{\rm Re} [(B_2 + D_2)(g C_{TE})^\ast]
    + 512 m_{B_c}^4 m_\ell \lambda \,
    \mbox{\rm Re} [(A_1 + C_1)(g C_T)^\ast] \nnb \\
&+&\frac{16}{3r} m_{B_c}^2 \Bigg( 4 (m_{B_c}^2 s + 8 m_\ell^2) \vel C_{TE} \ver^2
    + m_{B_c}^2 s v^2 \vel C_T \ver^2 \Bigg)
    \times \Bigg( 4 (\lambda+12 r s) \vel B_6 \ver^2 \nnb \\
&+& m_{B_c}^4 \lambda^2 \vel B_7 \ver^2
    - 4 m_{B_c}^2 (1-r-s) \lambda \,\mbox{\rm Re} (B_6 B_7^\ast)
    - 16 \, [\lambda + 12 r (1-r) ]\,\mbox{\rm Re} (g B_6^\ast) \nnb \\
&+& 8 m_{B_c}^2 (1+3r-s) \lambda \,\mbox{\rm Re} (g
    B_7^\ast)\Bigg)\nnb
    \Bigg\}~,
\eea
where $\lambda=1+r^2+s^2-2r-2s-2rs$ , $r=m_{D_{s}^\ast}^2/m_{B_c}^2$
and $v=\sqrt{1-\ds{\frac{4 m_\ell^2}{s m^2_{B_c}}}}$ is the lepton
velocity.

The polarizations $P^{\pm}_L$, $P^{\pm}_T$ and $P^{\pm}_N$ in Eq. (\ref{ddr}) are defined by the equation
\bea
P_i^{\pm}(q^2) = \frac{\ds{\frac{d \Gamma}{dq^2}
                   (\vec{n}^{\pm}=\vec{e}_i^{\,\pm}) -
                   \frac{d \Gamma}{dq^2}
                   (\vec{n}^{\pm}=-\vec{e}_i^{\,\pm})}}
              {\ds{\frac{d \Gamma}{dq^2}
                   (\vec{n}^{\pm}=\vec{e}_i^{\,\pm}) +
                  \frac{d \Gamma}{dq^2}
                  (\vec{n}^{\pm}=-\vec{e}_i^{\,\pm})}}~, \nnb
\eea
for $i=L,~N,~T$, i.e., $P^{\pm}_L$ and $P^{\pm}_T$ represents the charged lepton $\ell^{\pm}$
 longitudinal and transversal asymmetries in the decay plane,
respectively, and $P^{\pm}_N$ is the normal component to both of them.
After some lengthy algebra, we get for the longitudinal
polarization of the $\ell^{\pm}$
\bea \label{plm} P_L^{\pm}&=& \frac{4}{\Delta} m_{B_c}^2 v \Bigg\{\mp
    \frac{1}{3 r} \lambda^2 m_{B_c}^4 \Big[ \vel B_2 \ver^2 - \vel D_2
    \ver^2\Big] + \frac{1}{r} \lambda  m_\ell \,
    \mbox{\rm Re} [(B_1 - D_1) (B_4^\ast + B_5^\ast)] \nnb \\
&-& \frac{1}{r} \lambda m_{B_c}^2 m_\ell (1-r) \,
    \mbox{\rm Re} [(B_2 - D_2) (B_4^\ast + B_5^\ast)]\mp
    \frac{8}{3} \lambda m_{B_c}^4 s \Big[ \vel A_1 \ver^2 - \vel C_1 \ver^2\Big]
    \nnb \\
&-&\frac{1}{2r} \lambda m_{B_c}^2 s
    \Big[ \vel B_4 \ver^2 - \vel B_5 \ver^2\Big]-
    \frac{1}{r} \lambda m_{B_c}^2 m_\ell s \,
    \mbox{\rm Re} [(B_3 - D_3) (B_4^\ast + B_5^\ast)] \nnb \\
&\pm &\frac{2}{3 r} \lambda m_{B_c}^2 (1-r-s)
    \Big[ \mbox{\rm Re}(B_1 B_2^\ast) - \mbox{\rm Re}(D_1 D_2^\ast)\Big]
    \mp\frac{1}{3 r} (\lambda + 12 r s)
    \Big[ \vel B_1 \ver^2 - \vel D_1 \ver^2\Big] \nnb \\
&\mp&\frac{256}{3} \lambda m_{B_c}^2 m_\ell
    \,\Big( \mbox{\rm Re} [A_1^\ast (C_{T} \mp C_{TE})g] -
    \mbox{\rm Re} [C_1^\ast (C_T\pm C_{TE})g] \Big) \nnb \\
&+&\frac{4}{3r} \lambda^2 m_{B_c}^4 m_\ell \Big(
    \mbox{\rm Re} [B_2^\ast (C_T\mp 4 C_{TE}) B_7]
    + \mbox{\rm Re} [D_2^\ast (C_T\pm 4 C_{TE}) B_7] \Big)  \nnb \\
&-&\frac{8}{3r} \lambda m_{B_c}^2 m_\ell (1-r-s) \Big(
    \mbox{\rm Re} [B_2^\ast (C_T\mp 4 C_{TE}) B_6]
    +\mbox{\rm Re} [D_2^\ast (C_T\pm 4 C_{TE}) B_6] \Big) \nnb \\
&-&\frac{4}{3r} \lambda m_{B_c}^2 m_\ell (1-r-s) \Big(
    \mbox{\rm Re} [B_1^\ast (C_T\mp 4 C_{TE}) B_7]
    + \mbox{\rm Re} [D_1^\ast (C_T\pm 4 C_{TE}) B_7]  \Big) \nnb \\
&+&\frac{8}{3r} (\lambda+12 r s)  m_\ell \Big(
    \,\mbox{\rm Re} [B_1^\ast (C_T\mp 4 C_{TE}) B_6]
    +\mbox{\rm Re} [D_1^\ast (C_T\pm 4 C_{TE}) B_6] \Big) \nnb \\
&-&\frac{16}{3r}  m_\ell [\lambda+12 r (1-r)] \Big(
    \mbox{\rm Re} [B_1^\ast (C_T\mp 4 C_{TE}) g]
    +\mbox{\rm Re} [D_1^\ast (C_T\pm 4 C_{TE}) g] \Big)\\
&+&\frac{16}{3r} \lambda m_{B_c}^2 m_\ell (1+3 r - s) \Big(
    \mbox{\rm Re} [B_2^\ast (C_T\mp 4 C_{TE}) g]
    +\mbox{\rm Re} [D_2^\ast (C_T\pm 4 C_{TE}) g] \Big) \nnb \\
&+&\frac{16}{3r} \lambda^2 m_{B_c}^6 s
    \vel B_7 \ver^2 \mbox{\rm Re} (C_T C_{TE}^\ast) \nnb \\
&+&\frac{64}{3r} (\lambda + 12 r s) m_{B_c}^2 s
    \vel B_6 \ver^2 \mbox{\rm Re} (C_T C_{TE}^\ast) \nnb \\
&-&\frac{64}{3r} \lambda m_{B_c}^4 s (1-r-s)
    \,\mbox{\rm Re} (B_6 B_7^\ast) \mbox{\rm Re} (C_T C_{TE}^\ast) \nnb \\
&+&\frac{128}{3r} \lambda m_{B_c}^4 s (1+3 r-s)
    \,\mbox{\rm Re} (B_7 g^\ast) \mbox{\rm Re} (C_T C_{TE}^\ast) \nnb \\
&-&\frac{256}{3r} m_{B_c}^2 s [\lambda + 12 r (1-r)]
    \,\mbox{\rm Re} (B_6 g^\ast) \mbox{\rm Re} (C_T C_{TE}^\ast) \nnb \\
&+&\frac{256}{3r} m_{B_c}^2
    [ \lambda (4 r + s) + 12 r (1-r)^2 ] \vel g \ver^2
    \,\mbox{\rm Re} (C_T C_{TE}^\ast)
    \Bigg\}~,\nnb
\eea
where $\Delta$ is the term inside curly brackets of Eq. (\ref{unp}).

Similarly,  we find for the transverse polarization $P_T^{\pm}$
\bea
\label{ptm}
    P_T^-&=& \frac{\pi}{\Delta} m_{B_c} \sqrt{s \lambda} \Bigg\{
    -8 m_{B_c}^2 m_\ell  \, \mbox{\rm Re} [(A_1 + C_1) (B_1^\ast + D_1^\ast)] \nnb \\
&+& \frac{1}{r} m_{B_c}^2 m_\ell (1+3 r - s)  \,
    \Big[ \mbox{\rm Re}(B_1 D_2^\ast) -  \mbox{\rm Re}(B_2 D_1^\ast)\Big] \nnb \\
&+&\frac{1}{r s} m_\ell (1- r - s)
    \Big[ \vel B_1 \ver^2 - \vel D_1 \ver^2\Big] \nnb \\
&+&\frac{2}{r s} m_\ell^2 (1- r - s)
    \Big[ \mbox{\rm Re}(B_1 B_5^\ast) - \mbox{\rm Re}(D_1 B_4^\ast)\Big] \nnb \\
&-&\frac{1}{r} m_{B_c}^2 m_\ell (1- r - s) \,
    \mbox{\rm Re} [(B_1 + D_1) (B_3^\ast - D_3^\ast)] \nnb \\
&-&\frac{2}{r s} m_{B_c}^2 m_\ell^2 \lambda
    \Big[  \mbox{\rm Re}(B_2 B_5^\ast) -  \mbox{\rm Re}(D_2 B_4^\ast)\Big] \nnb \\
&+&\frac{1}{r s} m_{B_c}^4 m_\ell(1-r) \lambda
    \Big[ \vel B_2 \ver^2 - \vel D_2 \ver^2\Big]
    + \frac{1}{r} m_{B_c}^4 m_\ell \lambda  \,
    \mbox{\rm Re} [(B_2 + D_2) (B_3^\ast - D_3^\ast)] \nnb \\
&-&\frac{1}{r s} m_{B_c}^2 m_\ell [\lambda + (1-r-s) ( 1-r)]
    \Big[\mbox{\rm Re}(B_1 B_2^\ast) - \mbox{\rm Re}(D_1 D_2^\ast)\Big] \nnb \\
&+&\frac{1}{r s} (1-r-s)(2 m_\ell^2  - m_{B_c}^2 s )
    \Big[\mbox{\rm Re}(B_1 B_4^\ast)-\mbox{\rm Re}(D_1 B_5^\ast)\Big] \nnb \\
&+&\frac{1}{r s} m_{B_c}^2 \lambda (2 m_\ell^2  - m_{B_c}^2 s )
    \Big[\mbox{\rm Re}(D_2 B_5^\ast) - \mbox{\rm Re}(B_2 B_4^\ast)\Big] \nnb \\
&-&\frac{16}{r s} \lambda m_{B_c}^2 m_\ell^2
    \,\mbox{\rm Re}[(B_1-D_1) (B_7 C_{TE})^\ast] \nnb \\
&+&\frac{16}{r s} \lambda m_{B_c}^4 m_\ell^2 (1-r)
    \,\mbox{\rm Re}[(B_2-D_2) (B_7 C_{TE})^\ast] \nnb \\
&+&\frac{8}{r} \lambda m_{B_c}^4 m_\ell
    \,\mbox{\rm Re}[(B_4-B_5) (B_7 C_{TE})^\ast] \nnb \\
&+&\frac{16}{r} \lambda m_{B_c}^4 m_\ell^2
    \,\mbox{\rm Re}[(B_3-D_3) (B_7 C_{TE})^\ast] \nnb \\
&+&\frac{32}{r s} m_\ell^2 (1-r-s)
    \,\mbox{\rm Re}[(B_1-D_1) (B_6 C_{TE})^\ast] \\
&-&\frac{32}{r s} m_{B_c}^2 m_\ell^2 (1-r) (1-r-s)
    \,\mbox{\rm Re}[(B_2-D_2) (B_6 C_{TE})^\ast] \nnb \\
&-&\frac{16}{r} m_{B_c}^2 m_\ell  (1-r-s)
    \,\mbox{\rm Re}[(B_4-B_5) (B_6 C_{TE})^\ast] \nnb \\
&-&\frac{32}{r} m_{B_c}^2 m_\ell^2  (1-r-s)
    \,\mbox{\rm Re}[(B_3-D_3) (B_6 C_{TE})^\ast] \nnb \\
&-& 16 m_{B_c}^2  \Big(
    4 m_\ell^2 \, \mbox{\rm Re}[A_1^\ast (C_T+2 C_{TE}) B_6]
    - m_{B_c}^2 s \, \mbox{\rm Re}[A_1^\ast (C_T-2 C_{TE}) B_6] \Big)\nnb \\
&+& 16 m_{B_c}^2  \Big(
    4 m_\ell^2 \, \mbox{\rm Re}[C_1^\ast (C_T-2 C_{TE}) B_6]
    - m_{B_c}^2 s \, \mbox{\rm Re}[C_1^\ast (C_T+2 C_{TE}) B_6] \Big)\nnb \\
&+& \frac{32}{s} m_{B_c}^2 (1-r) \Big(
    4 m_\ell^2 \, \mbox{\rm Re}[A_1^\ast (C_T+2 C_{TE}) g]
    - m_{B_c}^2 s \, \mbox{\rm Re}[A_1^\ast (C_T-2 C_{TE}) g] \Big)\nnb \\
&-& \frac{32}{s} m_{B_c}^2 (1-r) \Big(
    4 m_\ell^2 \, \mbox{\rm Re}[C_1^\ast (C_T-2 C_{TE}) g]
    - m_{B_c}^2 s \, \mbox{\rm Re}[C_1^\ast (C_T+2 C_{TE}) g] \Big)\nnb \\
&+&\frac{64}{r s} m_{B_c}^2 m_\ell^2 (1-r) (1+3 r-s)
    \,\mbox{\rm Re}[(B_2- D_2) (g C_{TE})^\ast] \nnb \\
&+&\frac{64}{r} m_{B_c}^2 m_\ell^2  (1+3 r-s)
    \,\mbox{\rm Re}[(B_3-D_3) (g C_{TE})^\ast] \nnb \\
&+&\frac{32}{r} m_{B_c}^2 m_\ell  (1+3 r-s)
    \,\mbox{\rm Re}[(B_4-B_5) (g C_{TE})^\ast] \nnb \\
&+&\frac{64}{r s} [m_{B_c}^2 r s - m_\ell^2 (1+7 r -s)]
    \,\mbox{\rm Re}[(B_1-D_1) (g C_{TE})^\ast] \nnb \\
&-& \frac{32}{s} (4 m_\ell^2 + m_{B_c}^2 s)
    \,\mbox{\rm Re}[(B_1+D_1) (g C_T)^\ast] \nnb \\
&-&2048  m_{B_c}^2 m_\ell
    \,\mbox{\rm Re}[(C_T g)(B_6 C_{TE})^\ast] \nnb \\
&+&\frac{4096}{s} m_{B_c}^2 m_\ell (1-r) \vel g \ver^2
    \, \mbox{\rm Re}(C_T C_{TE}^\ast)
    \Bigg\}~, \nnb
\eea
and
\bea
\label{ptp}
    P_T^+&=& \frac{\pi}{\Delta} m_{B_c} \sqrt{s \lambda} \Bigg\{
    -8 m_{B_c}^2 m_\ell  \, \mbox{\rm Re} [(A_1 + C_1) (B_1^\ast + D_1^\ast)] \nnb \\
&-& \frac{1}{r} m_{B_c}^2 m_\ell (1+3 r - s)  \,
    \Big[ \mbox{\rm Re}(B_1 D_2^\ast) -  \mbox{\rm Re}(B_2 D_1^\ast)\Big] \nnb \\
&-&\frac{1}{r s} m_\ell (1- r - s)
    \Big[ \vel B_1 \ver^2 - \vel D_1 \ver^2\Big] \nnb \\
&+&\frac{1}{r s} (2 m_\ell^2  - m_{B_c}^2 s ) (1- r - s)
    \Big[ \mbox{\rm Re}(B_1 B_5^\ast) - \mbox{\rm Re}(D_1 B_4^\ast)\Big] \nnb \\
&+&\frac{1}{r} m_{B_c}^2 m_\ell (1- r - s) \,
    \mbox{\rm Re} [(B_1 + D_1) (B_3^\ast - D_3^\ast)] \nnb \\
&-&\frac{1}{r s} m_{B_c}^2 \lambda (2 m_\ell^2  - m_{B_c}^2 s )
    \Big[  \mbox{\rm Re}(B_2 B_5^\ast) -  \mbox{\rm Re}(D_2 B_4^\ast)\Big] \nnb \\
&-&\frac{1}{r s} m_{B_c}^4 m_\ell(1-r) \lambda
    \Big[ \vel B_2 \ver^2 - \vel D_2 \ver^2\Big]
    - \frac{1}{r} m_{B_c}^4 m_\ell \lambda  \,
    \mbox{\rm Re} [(B_2 + D_2) (B_3^\ast - D_3^\ast)] \nnb \\
&+&\frac{1}{r s} m_{B_c}^2 m_\ell [\lambda + (1-r-s) ( 1-r)]
    \Big[\mbox{\rm Re}(B_1 B_2^\ast) - \mbox{\rm Re}(D_1 D_2^\ast)\Big] \nnb \\
&+&\frac{2}{r s}  m_\ell^2 (1-r-s)
    \Big[\mbox{\rm Re}(B_1 B_4^\ast)-\mbox{\rm Re}(D_1 B_5^\ast)\Big] \nnb \\
&+&\frac{2}{r s} m_{B_c}^2  m_\ell^2 \lambda
    \Big[\mbox{\rm Re}(D_2 B_5^\ast) - \mbox{\rm Re}(B_2 B_4^\ast)\Big] \nnb \\
&+&\frac{16}{r s} \lambda m_{B_c}^2 m_\ell^2
    \,\mbox{\rm Re}[(B_1-D_1) (B_7 C_{TE})^\ast] \nnb \\
&-&\frac{16}{r s} \lambda m_{B_c}^4 m_\ell^2 (1-r)
    \,\mbox{\rm Re}[(B_2-D_2) (B_7 C_{TE})^\ast] \nnb \\
&-&\frac{8}{r} \lambda m_{B_c}^4 m_\ell
    \,\mbox{\rm Re}[(B_4-B_5) (B_7 C_{TE})^\ast] \nnb \\
&-&\frac{16}{r} \lambda m_{B_c}^4 m_\ell^2
    \,\mbox{\rm Re}[(B_3-D_3) (B_7 C_{TE})^\ast] \nnb \\
&-&\frac{32}{r s} m_\ell^2 (1-r-s)
    \,\mbox{\rm Re}[(B_1-D_1) (B_6 C_{TE})^\ast] \\
&+&\frac{32}{r s} m_{B_c}^2 m_\ell^2 (1-r) (1-r-s)
    \,\mbox{\rm Re}[(B_2-D_2) (B_6 C_{TE})^\ast] \nnb \\
&+&\frac{16}{r} m_{B_c}^2 m_\ell  (1-r-s)
    \,\mbox{\rm Re}[(B_4-B_5) (B_6 C_{TE})^\ast] \nnb \\
&+&\frac{32}{r} m_{B_c}^2 m_\ell^2  (1-r-s)
    \,\mbox{\rm Re}[(B_3-D_3) (B_6 C_{TE})^\ast] \nnb \\
&+& 16 m_{B_c}^2  \Big(
    4 m_\ell^2 \, \mbox{\rm Re}[A_1^\ast (C_T-2 C_{TE}) B_6]
    - m_{B_c}^2 s \, \mbox{\rm Re}[A_1^\ast (C_T+2 C_{TE}) B_6] \Big)\nnb \\
&-& 16 m_{B_c}^2  \Big(
    4 m_\ell^2 \, \mbox{\rm Re}[C_1^\ast (C_T+2 C_{TE}) B_6]
    - m_{B_c}^2 s \, \mbox{\rm Re}[C_1^\ast (C_T-2 C_{TE}) B_6] \Big)\nnb \\
&-& \frac{32}{s} m_{B_c}^2 (1-r) \Big(
    4 m_\ell^2 \, \mbox{\rm Re}[A_1^\ast (C_T-2 C_{TE}) g]
    - m_{B_c}^2 s \, \mbox{\rm Re}[A_1^\ast (C_T+2 C_{TE}) g] \Big)\nnb \\
&+& \frac{32}{s} m_{B_c}^2 (1-r) \Big(
    4 m_\ell^2 \, \mbox{\rm Re}[C_1^\ast (C_T+2 C_{TE}) g]
    - m_{B_c}^2 s \, \mbox{\rm Re}[C_1^\ast (C_T-2 C_{TE}) g] \Big)\nnb \\
&-&\frac{64}{r s} m_{B_c}^2 m_\ell^2 (1-r) (1+3 r-s)
    \,\mbox{\rm Re}[(B_2- D_2) (g C_{TE})^\ast] \nnb \\
&-&\frac{64}{r} m_{B_c}^2 m_\ell^2  (1+3 r-s)
    \,\mbox{\rm Re}[(B_3-D_3) (g C_{TE})^\ast] \nnb \\
&-&\frac{32}{r} m_{B_c}^2 m_\ell  (1+3 r-s)
    \,\mbox{\rm Re}[(B_4-B_5) (g C_{TE})^\ast] \nnb \\
&-&\frac{64}{r s} [m_{B_c}^2 r s - m_\ell^2 (1+7 r -s)]
    \,\mbox{\rm Re}[(B_1-D_1) (g C_{TE})^\ast] \nnb \\
&-& \frac{32}{s} (4 m_\ell^2 + m_{B_c}^2 s)
    \,\mbox{\rm Re}[(B_1+D_1) (g C_T)^\ast] \nnb \\
&-&2048  m_{B_c}^2 m_\ell
    \,\mbox{\rm Re}[(C_T g)(B_6 C_{TE})^\ast] \nnb \\
&+&\frac{4096}{s} m_{B_c}^2 m_\ell (1-r) \vel g \ver^2
    \, \mbox{\rm Re}(C_T C_{TE}^\ast)
    \Bigg\}~. \nnb
\eea

Finally for normal asymmetries we get

\bea
\label{pnm}
    P_N^-&=& \frac{1}{\Delta} \pi v m_{B_c}^3 \sqrt{s \lambda} \Bigg\{
    8 m_\ell \, \mbox{\rm Im}[(B_1^\ast C_1) + (A_1^\ast D_1)] \nnb \\
&-& \frac{1}{r} m_{B_c}^2 \lambda
    \,\mbox{\rm Im}[(B_2^\ast B_4) + (D_2^\ast B_5)] \nnb \\
&+& \frac{1}{r} m_{B_c}^2 m_\ell \lambda
    \,\mbox{\rm Im}[(B_2-D_2) (B_3^\ast-D_3^\ast)] \nnb \\
&-&\frac{1}{r} m_\ell \,(1 + 3 r - s)
    \,\mbox{\rm Im}[(B_1-D_1) (B_2^\ast-D_2^\ast)] \nnb \\
&+&\frac{1}{r} (1 - r - s)
    \, \mbox{\rm Im}[(B_1^\ast B_4) + (D_1^\ast B_5)] \nnb \\
&-&\frac{1}{r} m_\ell \,(1 - r - s)
    \,\mbox{\rm Im}[(B_1-D_1) (B_3^\ast-D_3^\ast)]  \nnb \\
&-&\frac{8}{r} m_{B_c}^2 m_\ell \lambda
    \,\mbox{\rm Im}[(B_4+B_5)(B_7 C_{TE})^\ast] \\
&+& \frac{16}{r} m_\ell \,(1-r-s)
    \,\mbox{\rm Im}[(B_4+B_5)(B_6 C_{TE})^\ast] \nnb \\
&-& \frac{32}{r} m_\ell \,(1+3 r-s)
    \,\mbox{\rm Im}[(B_4+B_5)(g C_{TE})^\ast] \nnb \\
&-& 16 m_{B_c}^2 s \Big(
    \,\mbox{\rm Im}[A_1^\ast (C_T-2 C_{TE}) B_6] +
    \mbox{\rm Im}[C_1^\ast (C_T+2 C_{TE}) B_6] \Big) \nnb \\
&+& 32 m_{B_c}^2 (1-r) \Big(
    \,\mbox{\rm Im}[A_1^\ast (C_T-2 C_{TE}) g] +
    \mbox{\rm Im}[C_1^\ast (C_T+2 C_{TE}) g] \Big) \nnb \\
&+& 32 \Big(
    \mbox{\rm Im}[B_1^\ast (C_T-2 C_{TE}) g]
    - \mbox{\rm Im}[D_1^\ast (C_T+2 C_{TE}) g] \Big) \nnb \\
&+& 512 m_\ell \,
    \ga \vel C_T \ver^2 - 4 \vel C_{TE} \ver^2 \dr
    \,\mbox{\rm Im}(B_6^\ast g)
    \Bigg\}~, \nnb
\eea
and
\bea
\label{pnp}
    P_N^+&=& \frac{1}{\Delta} \pi v m_{B_c}^3 \sqrt{s \lambda} \Bigg\{
    - 8 m_\ell \, \mbox{\rm Im}[(B_1^\ast C_1) + (A_1^\ast D_1)] \nnb \\
&+& \frac{1}{r} m_{B_c}^2 \lambda
    \,\mbox{\rm Im}[(B_2^\ast B_5) + (D_2^\ast B_4)] \nnb \\
&+& \frac{1}{r} m_{B_c}^2 m_\ell \lambda
    \,\mbox{\rm Im}[(B_2-D_2) (B_3^\ast-D_3^\ast)] \nnb \\
&-&\frac{1}{r} m_\ell \,(1 + 3 r - s)
    \,\mbox{\rm Im}[(B_1-D_1) (B_2^\ast-D_2^\ast)] \nnb \\
&-&\frac{1}{r} (1 - r - s)
    \, \mbox{\rm Im}[(B_1^\ast B_5) + (D_1^\ast B_4)] \nnb \\
&-&\frac{1}{r} m_\ell \,(1 - r - s)
    \,\mbox{\rm Im}[(B_1-D_1) (B_3^\ast-D_3^\ast)]  \nnb \\
&+&\frac{8}{r} m_{B_c}^2 m_\ell \lambda
    \,\mbox{\rm Im}[(B_4+B_5)(B_7 C_{TE})^\ast] \\
&-& \frac{16}{r} m_\ell \,(1-r-s)
    \,\mbox{\rm Im}[(B_4+B_5)(B_6 C_{TE})^\ast] \nnb \\
&+& \frac{32}{r} m_\ell \,(1+3 r-s)
    \,\mbox{\rm Im}[(B_4+B_5)(g C_{TE})^\ast] \nnb \\
&-& 16 m_{B_c}^2 s \Big(
    \,\mbox{\rm Im}[A_1^\ast (C_T+2 C_{TE}) B_6] +
    \mbox{\rm Im}[C_1^\ast (C_T-2 C_{TE}) B_6] \Big) \nnb \\
&+& 32 m_{B_c}^2 (1-r) \Big(
    \,\mbox{\rm Im}[A_1^\ast (C_T+2 C_{TE}) g] +
    \mbox{\rm Im}[C_1^\ast (C_T-2 C_{TE}) g] \Big) \nnb \\
&-& 32 \Big(
    \mbox{\rm Im}[B_1^\ast (C_T+2 C_{TE}) g]
    - \mbox{\rm Im}[D_1^\ast (C_T-2 C_{TE}) g] \Big) \nnb \\
&+& 512 m_\ell \,
    \ga \vel C_T \ver^2 - 4 \vel C_{TE} \ver^2 \dr
    \,\mbox{\rm Im}(B_6^\ast g)
    \Bigg\}~. \nnb
\eea

From Eqs. (\ref{plm})-(\ref{pnp}), we observe that for longitudinal and normal polarizations,
the difference between $\ell^+$ and $\ell^-$ lepton asymmetries results from the
scalar and tensor type interactions. Similar
situation takes place for transverse polarization asymmetries in the
$m_\ell \rar 0$ limit. From this, we can conclude that
 their experimental study may provide   essential information about new physics.

Another source of   useful information about new physics can be
a combined analysis of the lepton and antilepton polarizations, since in the SM $P_L^-+P_L^+=0$,
$P_N^-+P_N^+= 0$ and $P_T^- - P_T^+ \approx 0$ \cite{R19}.
Using Eqs. (\ref{plm})-(\ref{pnp}) we get

\bea
\label{lpl}
P_L^- + P_L^+ &=& \frac{4}{\Delta}  \,m_{B_c}^2 v \,\Bigg\{
    \frac{2}{r} m_\ell \lambda\,
    \mbox{\rm Re} [(B_1 - D_1) (B_4^\ast + B_5^\ast)] \nnb \\
&-& \frac{2}{r} m_{B_c}^2 m_\ell \lambda (1-r) \,
    \mbox{\rm Re} [(B_2 - D_2) (B_4^\ast + B_5^\ast)] \nnb \\
&-&\frac{1}{r} m_{B_c}^2 s \lambda
    \Big( \vel B_4 \ver^2 - \vel B_5 \ver^2\Big) -
    \frac{2}{r} m_{B_c}^2 m_\ell s \lambda \,
    \mbox{\rm Re} [(B_3 - D_3) (B_4^\ast + B_5^\ast)] \nnb \\
&+& \frac{8}{3 r} m_{B_c}^4 m_\ell \lambda^2 \,
    \mbox{\rm Re} [(B_2 + D_2) (B_7 C_T)^\ast] \nnb \\
&+& \frac{32}{3 r} m_{B_c}^6 s  \lambda^2 \vel B_7 \ver^2
    \mbox{\rm Re} (C_T C_{TE}^\ast) \nnb \\
&-& \frac{8}{3 r} m_{B_c}^2 m_\ell \lambda (1-r-s) \,
    \mbox{\rm Re} [(B_1 + D_1) (B_7 C_T)^\ast] \nnb \\
&-& \frac{16}{3 r} m_{B_c}^2 m_\ell \lambda (1-r-s) \,
    \mbox{\rm Re} [(B_2 + D_2) (B_6 C_T)^\ast] \nnb \\
&-& \frac{128}{3 r} m_{B_c}^4 s \lambda(1-r-s) \,
    \mbox{\rm Re} (B_6 B_7^\ast) \, \mbox{\rm Re} (C_T C_{TE}^\ast) \nnb \\
&+& \frac{16}{3 r}  m_\ell (\lambda + 12 r s) \,
    \mbox{\rm Re} [(B_1 + D_1) (B_6 C_T)^\ast] \\
&+& \frac{128}{3 r} m_{B_c}^2 s (\lambda + 12 r s) \,
    \vel B_6 \ver^2 \mbox{\rm Re} (C_T C_{TE}^\ast) \nnb \\
&+& \frac{512}{3 r} m_{B_c}^2 \, [ \lambda (4 r + s) + 12 r (1-r)^2 ]
    \, \vel g \ver^2 \,\mbox{\rm Re} (C_T C_{TE}^\ast)\nnb \\
&-& \frac{512}{3 r} m_{B_c}^2 s \, [\lambda +12 r (1-r) ]\,
    \mbox{\rm Re} (g B_6^\ast) \, \mbox{\rm Re} (C_T C_{TE}^\ast) \nnb \\
&+& \frac{256}{3 r} m_{B_c}^4 s \lambda (1+3r -s) \,
    \mbox{\rm Re} (g B_7^\ast) \, \mbox{\rm Re} (C_T C_{TE}^\ast) \nnb \\
&+& \frac{512}{3} m_{B_c}^2 m_\ell \lambda \,
    \mbox{\rm Re} [(A_1 + C_1) (g C_{TE})^\ast] \nnb \\
&-& \frac{32}{3 r}  m_\ell  \, [\lambda +12 r (1-r) ]\,
    \mbox{\rm Re} [(B_1 + D_1) (g C_T)^\ast] \nnb \\
&+& \frac{32}{3 r} m_{B_c}^2 m_\ell  \lambda (1+3r -s) \,
    \mbox{\rm Re} [(B_2 + D_2) (g C_T)^\ast]
    \Bigg\}~. \nnb
\eea

For the case of transverse  polarization, it is the difference of the lepton
and antilepton polarizations that is relevant and it can be calculated
from Eqs. (\ref{ptm}) and (\ref{ptp})
\bea
\label{tmt} P_T^- - P_T^+ &=& \frac{\pi}{\Delta} m_{B_c}
    \sqrt{s \lambda} \Bigg\{ \frac{2}{r s} m_{B_c}^4 m_\ell (1-r)
    \lambda
    \Big[ \vel B_2 \ver^2 - \vel D_2 \ver^2\Big]\nnb \\
&+& \frac{1}{r} m_{B_c}^4 \lambda \,
    \mbox{\rm Re} [(B_2 + D_2) (B_4^\ast - B_5^\ast)] \nnb \\
&+& \frac{2}{r} m_{B_c}^4 m_\ell \lambda \,
    \mbox{\rm Re} [(B_2 + D_2) (B_3^\ast - D_3^\ast)] \nnb \\
&+& \frac{2}{r} m_{B_c}^2 m_\ell (1+3 r - s)  \,
    \Big[ \mbox{\rm Re}(B_1 D_2^\ast) -  \mbox{\rm Re}(B_2 D_1^\ast)\Big] \nnb \\
&+& \frac{2}{rs} m_\ell (1-r-s)
    \Big[ \vel B_1 \ver^2 - \vel D_1 \ver^2\Big]\nnb \\
&-& \frac{1}{r} m_{B_c}^2 (1-r-s)
    \mbox{\rm Re} [(B_1 + D_1) (B_4^\ast - B_5^\ast)] \nnb \\
&-& \frac{2}{r} m_{B_c}^2 m_\ell (1-r-s)
    \mbox{\rm Re} [(B_1 + D_1) (B_3^\ast - D_3^\ast)] \nnb \\
&-&\frac{2}{r s} m_{B_c}^2 m_\ell [\lambda + (1-r) (1-r-s)]
    \Big[\mbox{\rm Re}(B_1 B_2^\ast) - \mbox{\rm Re}(D_1 D_2^\ast)\Big] \nnb \\
&-&\frac{32}{r s} m_{B_c}^2 m_\ell^2 \lambda \,
    \mbox{\rm Re}[(B_1-D_1) (B_7 C_{TE})^\ast] \nnb \\
&+&\frac{32}{r s} m_{B_c}^4 m_\ell^2 \lambda (1-r)\,
    \mbox{\rm Re}[(B_2-D_2) (B_7 C_{TE})^\ast] \\
&+&\frac{16}{r} m_{B_c}^4 m_\ell \lambda \,
    \mbox{\rm Re}[(B_4-B_5) (B_7 C_{TE})^\ast] \nnb \\
&+&\frac{32}{r} m_{B_c}^4 m_\ell^2 \lambda \,
    \mbox{\rm Re}[(B_3-D_3) (B_7 C_{TE})^\ast] \nnb \\
&+&\frac{64}{r s} m_\ell^2 (1-r-s) \,
    \mbox{\rm Re}[(B_1-D_1) (B_6 C_{TE})^\ast] \nnb \\
&-&\frac{64}{r s} m_{B_c}^2 m_\ell^2 (1-r)(1-r-s) \,
    \mbox{\rm Re}[(B_2-D_2) (B_6 C_{TE})^\ast] \nnb \\
&-&\frac{32}{r} m_{B_c}^2 m_\ell (1-r-s) \,
    \mbox{\rm Re}[(B_4-B_5) (B_6 C_{TE})^\ast] \nnb \\
&-&\frac{64}{r} m_{B_c}^2 m_\ell^2 (1-r-s) \,
    \mbox{\rm Re}[(B_3-D_3) (B_6 C_{TE})^\ast] \nnb \\
&+& 32 m_{B_c}^4 s v^2 \,
    \mbox{\rm Re}[(A_1-C_1) (B_6 C_T)^\ast] \nnb \\
&+&\frac{64}{r} m_{B_c}^2 m_\ell (1+3 r-s) \,
    \mbox{\rm Re}[(B_4-B_5) (g C_{TE})^\ast] \nnb \\
&-& 64 m_{B_c}^4 (1-r) v^2 \,
    \mbox{\rm Re}[(A_1-C_1) (g C_T)^\ast] \nnb \\
&+&\frac{128}{r s} [m_{B_c}^2 r s - m_\ell^2 (1+7 r-s)] \,
    \mbox{\rm Re}[(B_1-D_1) (g C_{TE})^\ast] \nnb \\
&+&\frac{128}{r s} m_{B_c}^2 m_\ell^2 (1-r) (1+3 r-s)
    \mbox{\rm Re}[(B_2-D_2) (g C_{TE})^\ast] \nnb \\
&+&\frac{128}{r} m_{B_c}^2 m_\ell^2 (1+3 r-s)
    \mbox{\rm Re}[(B_3-D_3) (g C_{TE})^\ast]
    \Bigg\}~. \nnb
\eea

In the same manner it follows from Eqs. (\ref{pnm}) and (\ref{pnp})
\bea
\label{npn}
    P_N^- + P_N^+ &=& \frac{1}{\Delta} \pi v m_{B_c}^3 \sqrt{s\lambda} \Bigg\{
    - \frac{2}{r} m_\ell (1+3 r -s) \,
    \mbox{\rm Im} [(B_1 - D_1) (B_2^\ast - D_2^\ast)] \nnb \\
&-& \frac{2}{r} m_\ell (1-r -s) \,
    \mbox{\rm Im} [(B_1 - D_1) (B_3^\ast - D_3^\ast)] \nnb \\
&-& \frac{1}{r} (1-r -s) \,
    \mbox{\rm Im} [(B_1 - D_1) (B_4^\ast - B_5^\ast)] \nnb \\
&+&\frac{2}{r} m_{B_c}^2  m_\ell \lambda \,
    \mbox{\rm Im} [(B_2 - D_2) (B_3^\ast - D_3^\ast)] \\
&+&\frac{1}{r} m_{B_c}^2 \lambda \,
    \mbox{\rm Im} [(B_2 - D_2) (B_4^\ast - B_5^\ast)] \nnb \\
&+& 32 m_{B_c}^2 s \, \mbox{\rm Im} [(A_1 + C_1)(B_6 C_T)^\ast] \nnb \\
&+&1024 m_\ell \Big(  \vel C_T \ver^2  - \vel 4 C_{TE} \ver^2  \Big)
    \mbox{\rm Im} (B_6^\ast g) \nnb \\
&-& 64 m_{B_c}^2 (1-r) \, \mbox{\rm Im} [(A_1 + C_1)(g C_T)^\ast] \nnb \\
&+& 128 \, \mbox{\rm Im} [(B_1 + D_1)(g C_{TE})^\ast]
    \Bigg\}~. \nnb
\eea
It can be seen  from Eq. (\ref{lpl}) that in $P_L^-+P_L^+$ the terms containing the  SM contribution,
i.e., terms containing $C_{BR}$, $C_{SL}$, $C^{tot}_{LL}$ and $C^{tot}_{LR}$  completely cancels.
For this reason, a measurement of the nonzero value of $P_L^-+P_L^+$ in future experiments, may be  an indication of the
discovery of new physics beyond the SM.

Before going into the details of our numerical analysis we like to note a final point about the numerical calculations
of the polarization asymmetries. As seen from the Eqs.(\ref{plm})-(\ref{npn}),
all  the expressions of the lepton
polarizations    depend on both $s=q^2/m^2_{B_c}$ and the new Wilson coefficients. However,  it may be experimentally
easier to study the dependence of the
 polarizations of each lepton  on the new Wilson coefficients only. For this reason we
 eliminate $s$ dependence by considering their   averaged forms
over the allowed kinematical region.  The averaged lepton
polarizations are defined as \bea \label{av} \lla P_i \rra =
\frac{\ds \int_{(2 m_\ell/m_{B_c})^2}^{(1-m_{D_{s}^\ast}/m_{B_c})^2}
P_i \frac{d{\cal B}}{ds} ds} {\ds \int_{(2
m_\ell/m_{B_c})^2}^{(1-m_{D_{s}^\ast}/m_{B_c})^2}
 \frac{d{\cal B}}{ds} ds}~.
\eea
%
\section{Numerical analysis and discussion}
We here present  our numerical analysis about the branching ratios and averaged polarization asymmetries
 $<P^-_L>$, $<P^-_T>$
and $<P^-_N>$ of $\ell^-$ for the $B_c \rar D^* \ell^+ \ell^- $
decays with $\ell =\mu , \tau $, as well as the lepton-antilepton
combined asymmetries $<P^-_L+P^+_L>$, $<P^-_T-P^+_T>$ and
$<P^-_N+P^+_N>$. We first give the input parameters used in our
numerical analysis :
\begin{eqnarray}
\label{parameters}
 & & m_{B_c} =6.50 \, GeV \, , \,m_{D_{s}^\ast}=2.112 \,GeV\,\,\, m_b =4.8 \, GeV \, ,
\,m_{\mu} =0.105 \, GeV \, , \,
m_{\tau} =1.77 \, GeV \, , \nnb \\
& &  |V_{tb} V^*_{ts}|=0.0385 \, \, , \, \, \alpha^{-1}=129  \, \,  ,
G_F=1.17 \times 10^{-5}\, GeV^{-2} \nnb \\
& &  \tau_{B_{c}}=0.46 \times 10^{-12} \, s \, .
\end{eqnarray}
The values of the individual Wilson coefficients that appear in
the SM are listed in Table (\ref{table1}). The values for the mass
and the lifetime of the $B_c$ meson given above in
Eq.(\ref{parameters}) were reported by CDF Collaboration
\cite{CDFI}. Recently, CDF quoted a new value $m_{B_c} =6.2857 \pm
0.0053\pm 0.0012$ \, GeV \cite{CDFII}. Also, D0 has observed $B_c$
and reported the preliminary results $m_{B_c} =5.95^{+0.14}_{-0.13} \pm 0.34$ GeV
 and $\tau_{B_{c}}=0.45^{+0.12}_{-0.10} \pm 0.12$ \cite{D0}. However, we observed that our
 numerical results are not sensitive to the numerical values of $m_{B_c}$  more than $3-5\%$.

\begin{table}
        \begin{center}
        \begin{tabular}{|c|c|c|c|c|c|c|c|c|}
        \hline
        \multicolumn{1}{|c|}{ $C_1$}       &
        \multicolumn{1}{|c|}{ $C_2$}       &
        \multicolumn{1}{|c|}{ $C_3$}       &
        \multicolumn{1}{|c|}{ $C_4$}       &
        \multicolumn{1}{|c|}{ $C_5$}       &
        \multicolumn{1}{|c|}{ $C_6$}       &
        \multicolumn{1}{|c|}{ $C_7^{\rm eff}$}       &
        \multicolumn{1}{|c|}{ $C_9$}       &
                \multicolumn{1}{|c|}{$C_{10}$}      \\
        \hline
        $-0.248$ & $+1.107$ & $+0.011$ & $-0.026$ & $+0.007$ & $-0.031$ &
   $-0.313$ &   $+4.344$ &    $-4.624$       \\
        \hline
        \end{tabular}
        \end{center}
\caption{ Values of the SM Wilson coefficients at $\mu \sim m_b $ scale.\label{table1}}
\end{table}

We   note  that the value of the Wilson coefficient $C^{eff}_9$ in Table (\ref{table1})
 corresponds
only to the short-distance contributions. $C^{eff}_9$ also receives long-distance
contributions due to conversion of the real $\bar{c}c$ into
lepton pair $\ell^+ \ell^-$ and they are usually absorbed into a redefinition of the short-distance Wilson
coefficients:
\begin{eqnarray}
C_9^{eff}(\mu)=C_9(\mu)+ Y(\mu)\,\, ,
\label{C9efftot}
\end{eqnarray}
where
\begin{eqnarray}
\label{EqY}
Y(\mu)&=& Y_{reson}+ h(y,s) [ 3 C_1(\mu) + C_2(\mu) +
3 C_3(\mu) + C_4(\mu) + 3 C_5(\mu) + C_6(\mu)] \nonumber \\&-&
\frac{1}{2} h(1, s) \left( 4 C_3(\mu) + 4 C_4(\mu)
+ 3 C_5(\mu) + C_6(\mu) \right)\nnb \\
&- &  \frac{1}{2} h(0,  s) \left[ C_3(\mu) + 3 C_4(\mu) \right]
\\&+& \frac{2}{9} \left( 3 C_3(\mu) + C_4(\mu) + 3 C_5(\mu) +
C_6(\mu) \right) \nonumber \,\, ,
\end{eqnarray}
with $y=m_c/m_b$, and the functions $h(y,s)$ arises from
the one loop contributions of the four quark operators $O_1$,...,$O_6$ and their explicit
forms can be found in \cite{Misiak}.
It is possible to parametrize  the resonance $\bar{c}c$
contribution $Y_{reson}(s)$ in Eq.(\ref{EqY}) using a Breit-Wigner
shape with normalizations fixed by data which is given by
\cite{AAli2}
\begin{eqnarray}
Y_{reson}(s)&=&-\frac{3}{\alpha^2_{em}}\kappa \sum_{V_i=\psi_i}
\frac{\pi \Gamma(V_i\rightarrow \ell^+
\ell^-)m_{V_i}}{s m^2_{B_{c}}-m_{V_i}+i m_{V_i}
\Gamma_{V_i}} \nonumber \\
&\times & [ (3 C_1(\mu) + C_2(\mu) + 3 C_3(\mu) + C_4(\mu) + 3
C_5(\mu) + C_6(\mu))]\, ,
 \label{Yresx}
\end{eqnarray}
where the phenomenological parameter $\kappa$ is usually taken as
$\sim 2.3$.

As for the values of the new Wilson coefficients, they are the free parameters in this work,
but  it is possible to establish ranges out of experimentally measured branching ratios of the
semileptonic and also purely leptonic rare B-meson decays
\bea
BR (B \rar K \, \ell^+ \ell^-) & = & (0.75^{+0.25}_{-0.21}\pm 0.09) \times 10^{-6} \, \, ,\nnb \\
BR (B \rar K^* \, \mu^+ \mu^-) & = & (0.9 ^{+1.3}_{-0.9}\pm 0.1)\times 10^{-6}\, \, ,\nnb
\eea
reported  by Belle and Babar collaborations \cite{ABE}. It is now also available an upper bound of pure
leptonic rare B-decays in the $B^0 \rar  \mu^+ \mu^-$ mode \cite{Halyo}:
\bea
BR ( B^0 \rar  \mu^+ \mu^-) & \leq & 2.0 \times 10^{-7}  \, \, .\nnb
\eea
Being in accordance with this upper limit and also the above mentioned measurements of the branching ratios
for the semileptonic rare B-decays, we take in this work all new Wilson coefficients as real and varying in
the region $-4\leq C_X\leq 4$.

Among the new Wilson coefficients that appear in Eq.(\ref{effH}), those related to the
helicity-flipped counter-parts of the SM operators, namely, $C_{RL}$ and $C_{RR}$, vanish in all models with
minimal flavor violation in the limit $m_s \rar 0$. However, there are some MSSM scenarios in which
there are finite contributions from these vector operators even for a vanishing s-quark mass. In addition,
scalar type interactions can also contribute through the neutral Higgs diagrams in e.g. multi-Higgs doublet models
and MSSM  for some regions of the parameter spaces of the related models.
In literature there exists studies to establish ranges out of constraints under various precision measurements
for these coefficients (see e.g. \cite{HuangWu}) and our choice for the range of the new Wilson coefficients are
in agreement with these calculations.

To make some numerical predictions, we also need the explicit forms
of the form factors $A_{0}, A_{+}, A_{-}, V, a_{0}, a_{+}$ and $g$.
In our work we have used the results of \cite{Faessler}, in which
 $q^2$ dependencies of the form factors
are given as
\begin{eqnarray}
F(q^2) = \frac{F(0)}{\Big( 1-a s + b s^2 \Big)^2}~, \nnb
\end{eqnarray}
where the values of parameters $F(0)$, $a$ and $b$ for the $B_c \rar
D_{s}^\ast$ decay are listed in Table 2.
\begin{table}[h]
\renewcommand{\arraystretch}{1.5}
\addtolength{\arraycolsep}{3pt}
$$
\begin{array}{|l|ccc|}
\hline
& F(0) & a & b \\ \hline
A_0 & \phantom{-}0.279  & 1.30 & 0.149 \\
A_+ & \phantom{-}0.156  & 2.16 & \phantom{-}1.15\\
A_- & \phantom{-}-0.321  & 2.41 & \phantom{-}1.51\\
V   & \phantom{-}0.290  & 2.40 & \phantom{-}1.49\\
a_0 & \phantom{-}0.178  & 1.21 & \phantom{-}0.125\\
a_+ & \phantom{-}0.178  & 2.14 & 1.14\\
g   & \phantom{-}0.179  & 2.51 & \phantom{-}1.67 \\ \hline
\end{array}
$$
\caption{$B_c$ meson decay form factors in a relativistic constituent quark model.}
\renewcommand{\arraystretch}{1}
\addtolength{\arraycolsep}{-3pt}
\end{table}

We present the results of our analysis  in a series of figures.
Before the discussion of these figures, we give our SM predictions
for the longitudinal, transverse and the normal components of the
lepton polarizations for \BcDll decay for $\mu$ ($\tau$) channel
for reference: \bea
<P^-_{L}>  & =  & 0.6211 \, (0.6321) \, ,\nnb \\
<P^-_{T}>  & =  & 0.0017 \, (0.0468) \, ,\nnb \\
<P^-_{N}>  & =  & -0.0837 \, (-0.17) \, .\nnb
\eea

Figs. (\ref{f1}) and (\ref{f2}) give dependence of the integrated
branching ratio (BR) on the new Wilson coefficients for the $B_c
\rar D_{s}^\ast \, \mu^+ \mu^-$  and $B_c \rar D_{s}^\ast \, \tau^+
\tau^-$ decays, respectively. From these figures we see that BR
depends strongly on the tensor interactions and weakly on the vector
interactions, while it is completely insensitive to the scalar type
of interactions. It is also clear from these figures that dependence
of the BR on the new Wilson coefficients is symmetric with respect
to the zero point for the muon final state, but such a symmetry is
not observed for the tau final state for the tensor interactions.

In Figs. (\ref{f3}) and (\ref{f4}), we present the dependence of
averaged longitudinal polarization $<P_L^->$ of $\ell^-$ and the
combined averaged $<P_L^- + P_L^+ >$ for $B_c \rar D_{s}^\ast \mu^+
\mu^- $ decay on the new Wilson coefficients. We observe that
$<P_L^- >$ is  more sensitive to the existence of the tensor type
interactions while the combined average $<P_L^- + P_L^+ >$ is to
that of scalar type interactions only. The fact that $<P_L^- + P_L^+
>$ does not exhibit any dependence on the vector type of
interactions are already an expected result since vector type
interactions are cancelled when the longitudinal polarization
asymmetry of the lepton and antilepton is considered together. We
also note that the values of $<P_L^- >$ becomes substantially
different from the SM value (at $C_X=0$) as $C_X$ becomes different
from zero, which indicates that measurement of the longitudinal
lepton polarization in  $B_c \rar D_{s}^\ast \mu^+ \mu^- $ decay can
be very useful to investigate new physics beyond the SM. From Fig.
({\ref{f3}), we see that, the contributions coming from all types of
interactions to $<P_L^- >$ are positive and it is an increasing
(decreasing) functions of  both $C_T$ and $C_{TE}$ for their
negative (positive) values.  We  observe from Fig. (\ref{f4}) that
$<P_L^- + P_L^+ >$ becomes zero at $C_X=0$, which conforms the SM
results, and its dependence on $C_X$ is symmetric with respect to
this zero value. It is also interesting to note  that $<P_L^- +
P_L^+ >$ is positive for all values of $C_{LRLR}$ and $C_{RLLR}$,
while it is negative for remaining scalar type interactions.

Figs. (\ref{f5}) and (\ref{f6}) are the same as Figs. (\ref{f3}) and
(\ref{f4}), but for $B_c \rar D_{s}^\ast \tau^+ \tau^- $. Similar to
the muon case, $<P_L^- >$ is more  sensitive to the tensor
interactions than others. Contributions to $<P_L^- >$  from all type
of interactions are positive for all values of $C_X$ except for
$C_{TE}$: in region $0.25\simlt C_{TE}<4$, $<P_L^- >$ changes the
sign and becomes
 negative. As for the main interesting point in Fig. (\ref{f6}),
although $<P_L^- + P_L^+ >$ for $B_c \rar D_{s}^\ast \mu^+ \mu^- $
decay depends only on scalar interactions, for $B_c \rar D_{s}^\ast
\tau^+ \tau^- $ decay it is also and very sensitively dependent on
tensor type of interactions. It is also interesting to note that
$<P_L^- + P_L^+ >$ changes sign: it takes positive (negative) values
for the negative (positive) values of  $C_T$ and $C_{TE}$. Thus, one
can provide valuable information about the new physics by
determining  the sign and the magnitude of $<P_L^- + P_L^+ >$. We
finally note that as in case of muon final state, in tau final state
too, $<P_L^- + P_L^+ >$  becomes zero at $C_X=0$ and confirms the SM
result.

In Figs. (\ref{f7}) and (\ref{f8}), we present the dependence of
averaged transverse polarization $<P_T^->$ of $\ell^-$ and the
combined averaged $<P_T^- - P_T^+ >$ for $B_c \rar D_{s}^\ast \mu^+
\mu^- $ decay on the new Wilson coefficients. From these figures, it
is seen that for the $<P_T^- >$, there appears strong dependence on
tensor and scaler interactions and also a weak dependence on vector
interactions. On the other hand, vector contributions to the $<P_T^-
- P_T^+ >$ is negligible and main contribution comes from the tensor
interactions and $C_{LRRL}$ and $C_{RLRL}$ components of the scalar
interactions.
 As seen from Figs.
(\ref{f7}) and (\ref{f8}), both $<P_T^->$ and $<P_T^- - P_T^+ >$  are positive
(negative) for the negative (positive) values of $C_T$ and $C_{RLRL}$, except in a region about the zero
values of the coefficients,  $-1 \simlt C_X \simlt 1$, while their
behavior with respect to $C_{TE}$ and $C_{LRRL}$ are opposite.
Therefore, determination of the
sign and magnitude of these observables can also give useful information
about existence of new physics.

Figs. (\ref{f9}) and (\ref{f10}) are the same as Figs. (\ref{f7})
and (\ref{f8}), but for $B_c \rar D_{s}^\ast \tau^+ \tau^- $. We see
from Fig. (\ref{f9}) that the $<P_T^- >$ is quite sensitive to all
types of interactions and  behavior of scalar interaction is
identical for coefficients $C_{LRRL}, C_{LRLR}$ and $C_{RLLR},
C_{RLRL}$ in pairs. It can be seen from Fig. (\ref{f10}) that
although tensor and scalar interactions are dominant for $<P_T^- -
P_T^+ >$,  the dependence of vector interactions are also more
sizable as compared with the case of muon final state. In addition,
change in sign of $<P_T^- >$ and $<P_T^- - P_T^+ >$ are observed
depending on the change in the tensor and scalar interaction
coefficients, whose measure may provide useful tools for new
physics.

In Figs.(\ref{f11}) and (\ref{f12}), we present the dependence of
averaged normal polarization $<P_N^->$ of $\ell^-$ and the combined
averaged $<P_N^- + P_N^+ >$ for $B_c \rar D_{s}^\ast \mu^+ \mu^- $
decay on the new Wilson coefficients. We see from Fig.  (\ref{f11})
that $<P_N^->$ strongly depends on the tensor  interactions. Its
dependence on the scalar type of interactions is moderate and
identical for the coefficients $C_{LRRL}, C_{LRLR}$ and $C_{RLLR},
C_{RLRL}$ in pairs. As seen from Fig. (\ref{f12}), the behavior of
$<P_N^- + P_N^+ >$ is determined by  the tensor interactions only.
We also observe that   $<P_N^- + P_N^+ >$ is positive (negative)
when $C_T<0$ ($C_T>0$) while  its behavior with respect to $C_{TE}$
is  opposite. Further, $<P_N^- + P_N^+>$ becomes zero at $C_X=0$ as
expected in the SM.

Figs. (\ref{f13}) and (\ref{f14}) are the same as Figs. (\ref{f11})
and (\ref{f12}), but for $B_c \rar D_{s}^\ast \tau^+ \tau^- $. We
first note that as being  opposite to the muon final state case,
here $<P_N^->$ depends on all types of interactions although
dependence on tensor interaction  is stronger. We also observe that
$<P_N^->$ always takes the positive values except  when
$C_{TE}\simlt -0.25$  and $C_T \simgt 2$. As seen from Fig.
(\ref{f14}), $<P_N^- + P_N^+>$ depends only on the tensor
interactions and its behavior is  the same as that of the muon final
state case.
 \\

In conclusion, we present the most general analysis of the lepton
polarization asymmetries in the rare \BcDll decay using the general,
model independent form of the effective Hamiltonian. The dependence
of the longitudinal, transversal and normal polarization asymmetries
of $\ell^-$
 and their combined asymmetries on the new Wilson
coefficients are studied. It is found that the lepton polarization
asymmetries are very sensitive to the existence of the tensor and
scalar type interactions. Moreover, $<P_T >$ and $<P_N >$ change
their signs  as the new Wilson coefficients vary in the region of
interest. This conclusion is valid also for the combined
polarization effects $<P_L^- + P_L^+ >,~<P_T^- - P_T^+ >$ and
$<P_N^- + P_N^+ >$ for the same decay channel. It is well known that
in the SM, $<P_L^- + P_L^+> = <P_T^- - P_T^+> = <P_N^- + P_N^+>
\simeq 0$ in the limit $m_\ell \rar 0$. Therefore any deviation from
this relation and determination of the sign of polarization is
decisive and effective tool in looking for new physics beyond the
SM. \\
\\
{\large \bf Acknowledgments}\\
\\
This work was partially supported by Mersin University under Grant
No: BAP-FEF-FB (UOY) 2006-3.

\newpage
\renewcommand{\topfraction}{.99}
\renewcommand{\bottomfraction}{.99}
\renewcommand{\textfraction}{.01}
\renewcommand{\floatpagefraction}{.99}

\begin{figure}
\centering
\includegraphics[width=5in]{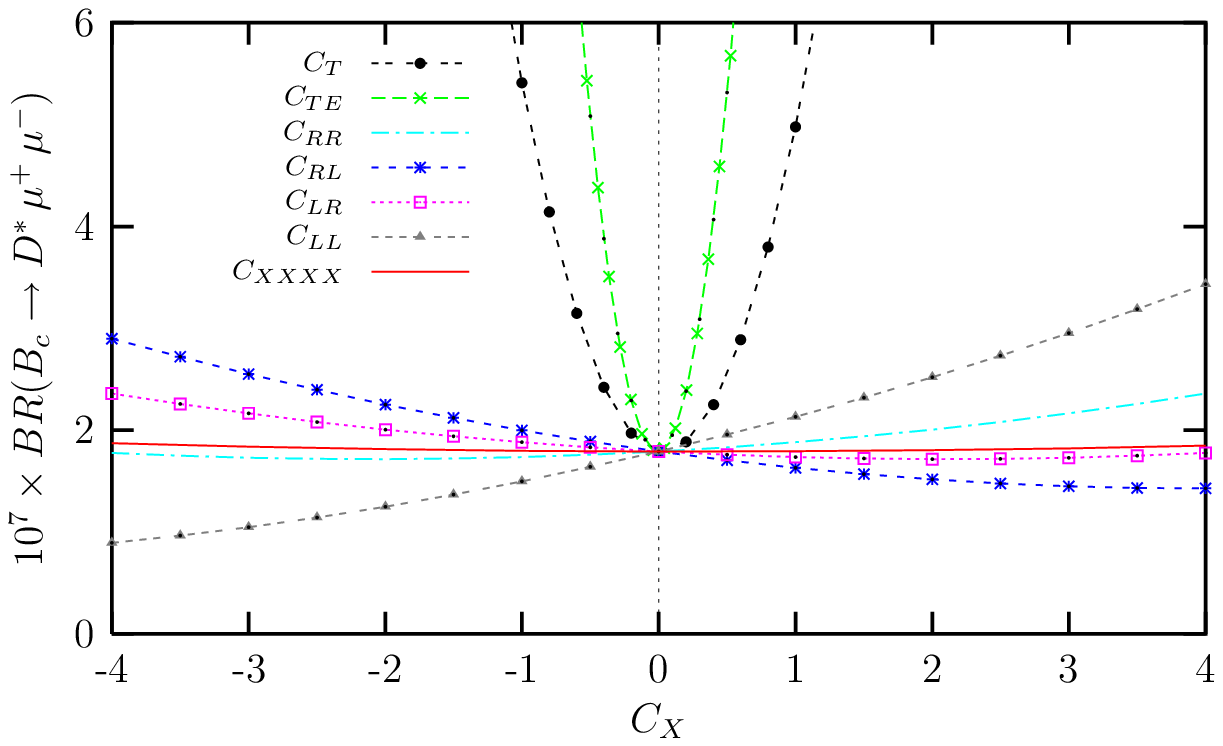}
\caption{The dependence of the integrated branching ratio for the
$B_c \rar D_{s}^{\ast} \, \mu^+ \mu^-$ decay on the new Wilson
coefficients. \label{f1}}
\end{figure}
\begin{figure}
\centering
\includegraphics[width=5in]{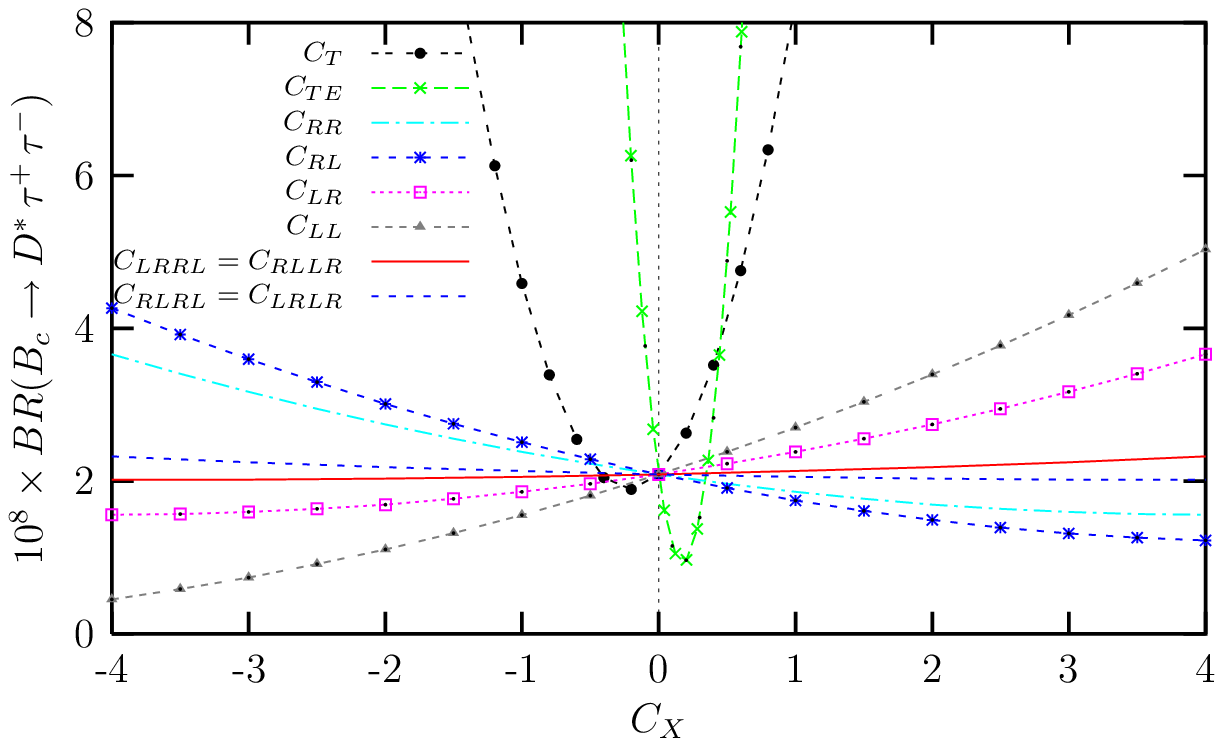}
\caption{The dependence of the integrated branching ratio for the
$B_c \rar D_{s}^{\ast} \, \tau^+ \tau^-$ decay on the new Wilson
coefficients. \label{f2}}
\end{figure}
\clearpage
\begin{figure}
\centering
\includegraphics[width=5in]{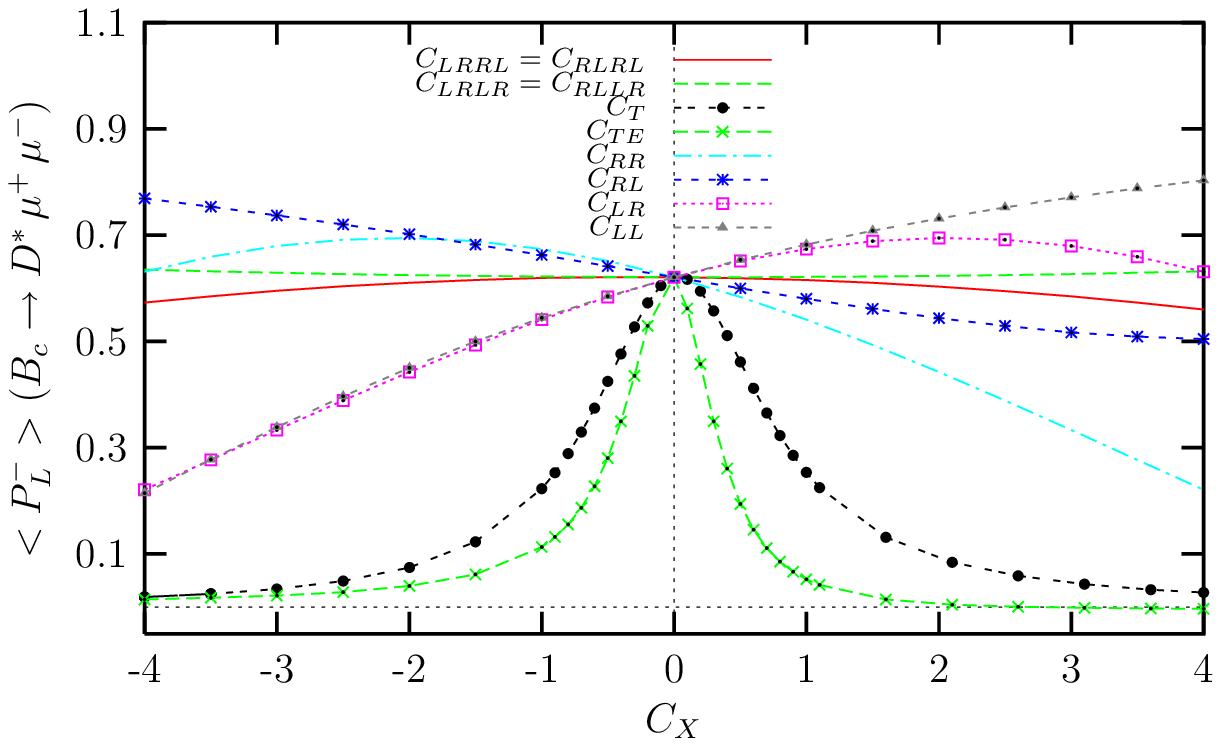}
\caption{The dependence of the averaged longitudinal polarization
$<P^-_L>$ of $\ell^-$ for the $B_c \rar D_{s}^{\ast}\, \mu^+ \mu^-$
decay on the new Wilson coefficients. \label{f3}}
\end{figure}
\begin{figure}
\centering
\includegraphics[width=5in]{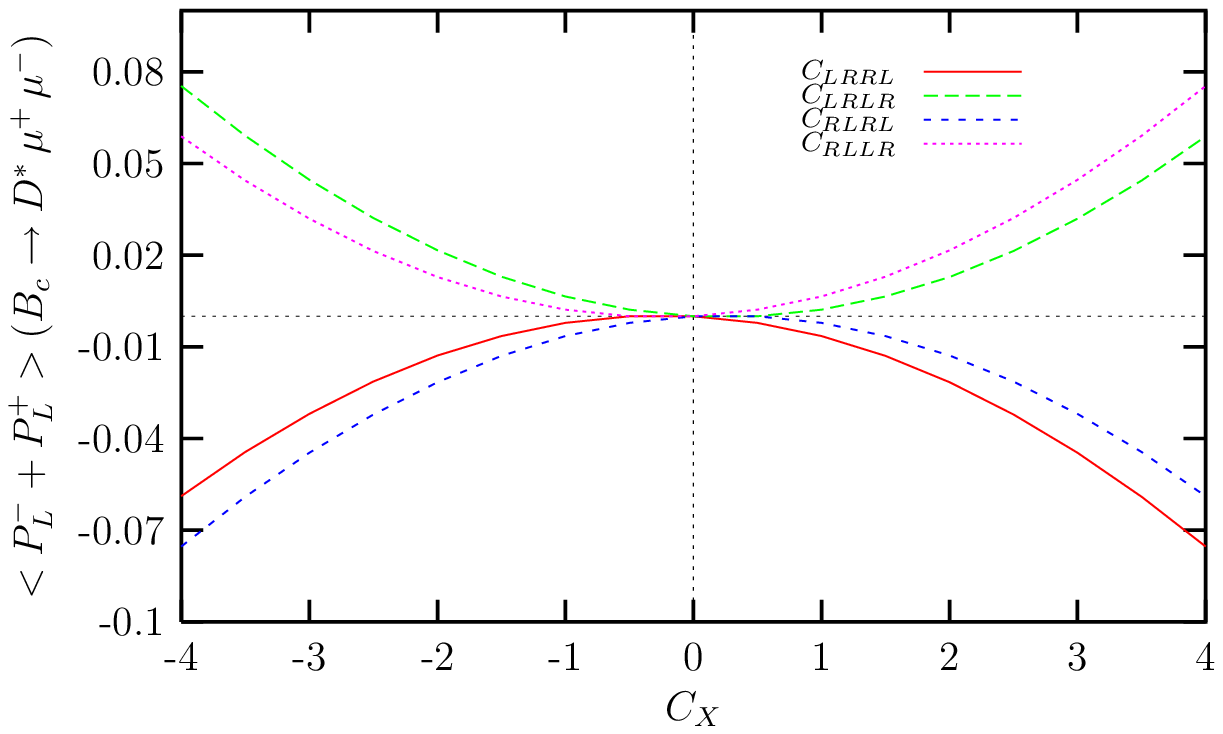}
\caption{The dependence of the combined averaged longitudinal lepton
polarization \, \, \, $<P^-_L+P^+_L>$ for the $B_c \rar D_{s}^{\ast}
\, \mu^+ \mu^-$  decay on the new Wilson coefficients.\label{f4}}
\end{figure}
\clearpage
\begin{figure}
\centering
\includegraphics[width=5in]{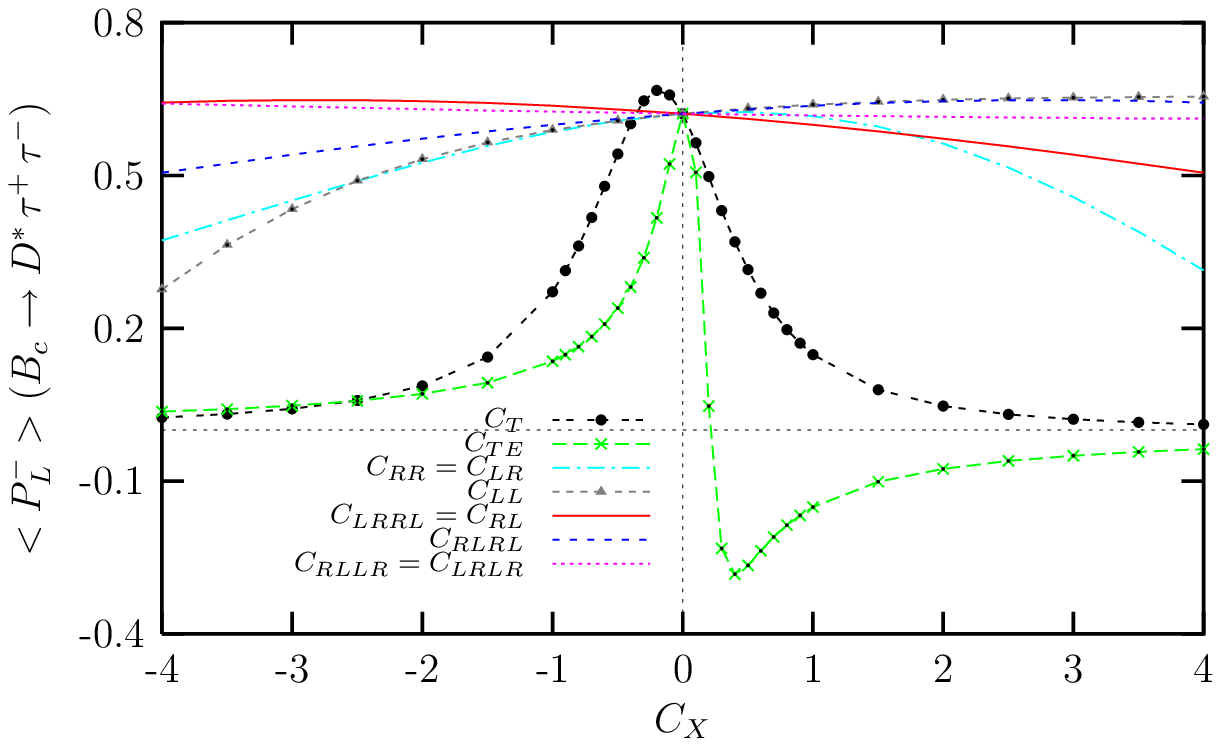}
\caption{The same as Fig. (\ref{f3}), but for the $B_c \rar
 D_{s}^{\ast}\, \tau^+ \tau^-$ decay. \label{f5}}
\end{figure}
\begin{figure}
\centering
\includegraphics[width=5in]{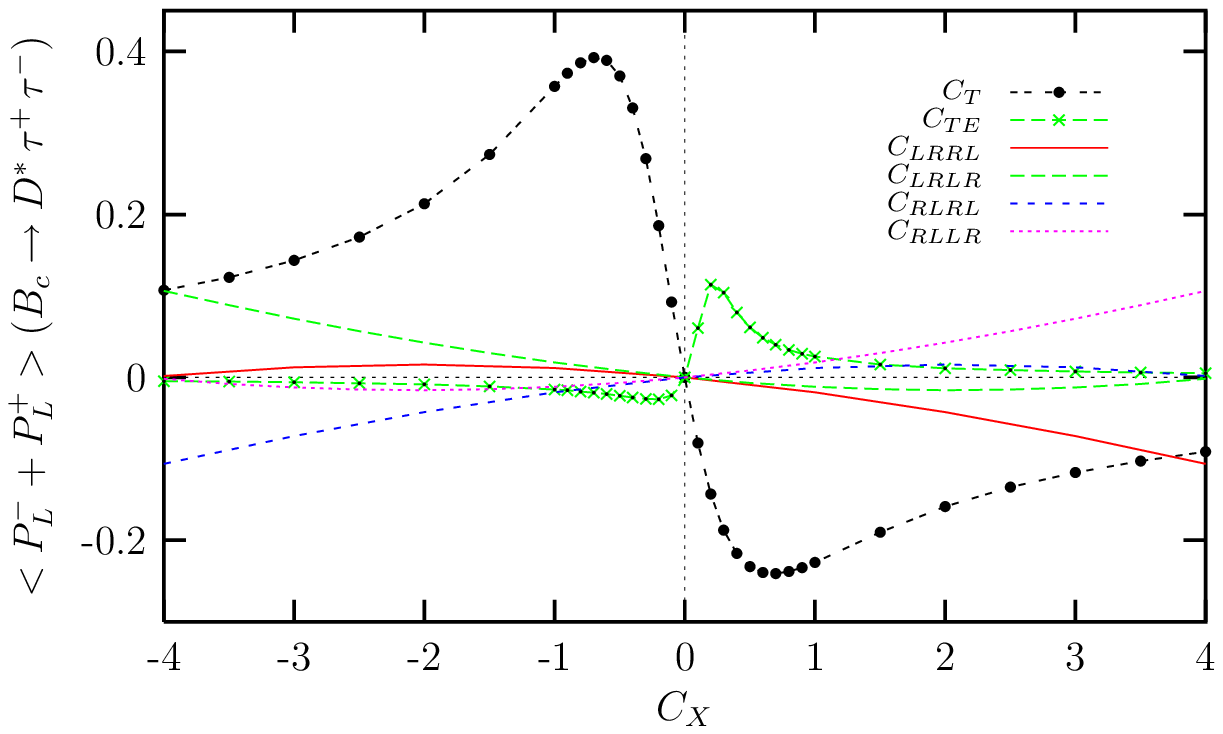}
\caption{The same as Fig. (\ref{f4}), but for the $B_c \rar
 D_{s}^{\ast}\, \tau^+ \tau^-$ decay. \label{f6}}
\end{figure}
\clearpage
\begin{figure}
\centering
\includegraphics[width=5in]{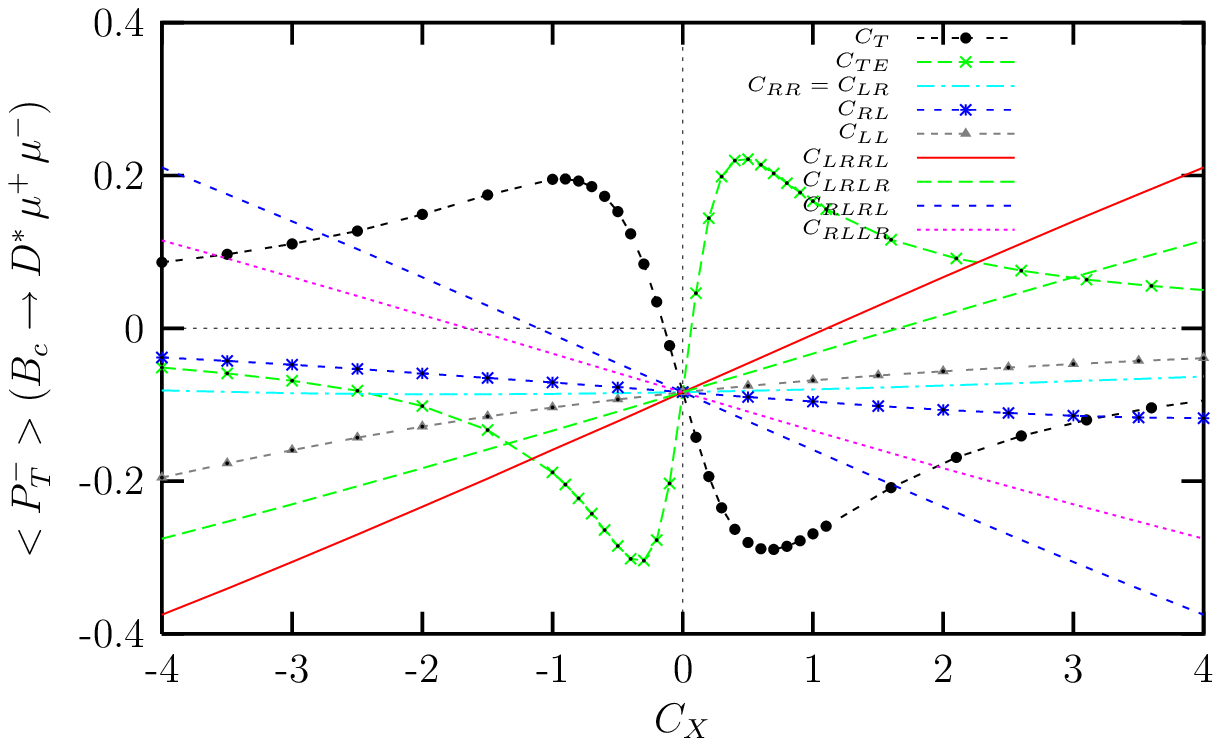}
\caption{The dependence of the averaged transverse polarization
$<P^-_T>$ of $\ell^-$ for the $B_c \rar D_{s}^{\ast} \, \mu^+ \mu^-$
decay on the new Wilson coefficients. \label{f7}}
\end{figure}
\begin{figure}
\centering
\includegraphics[width=5in]{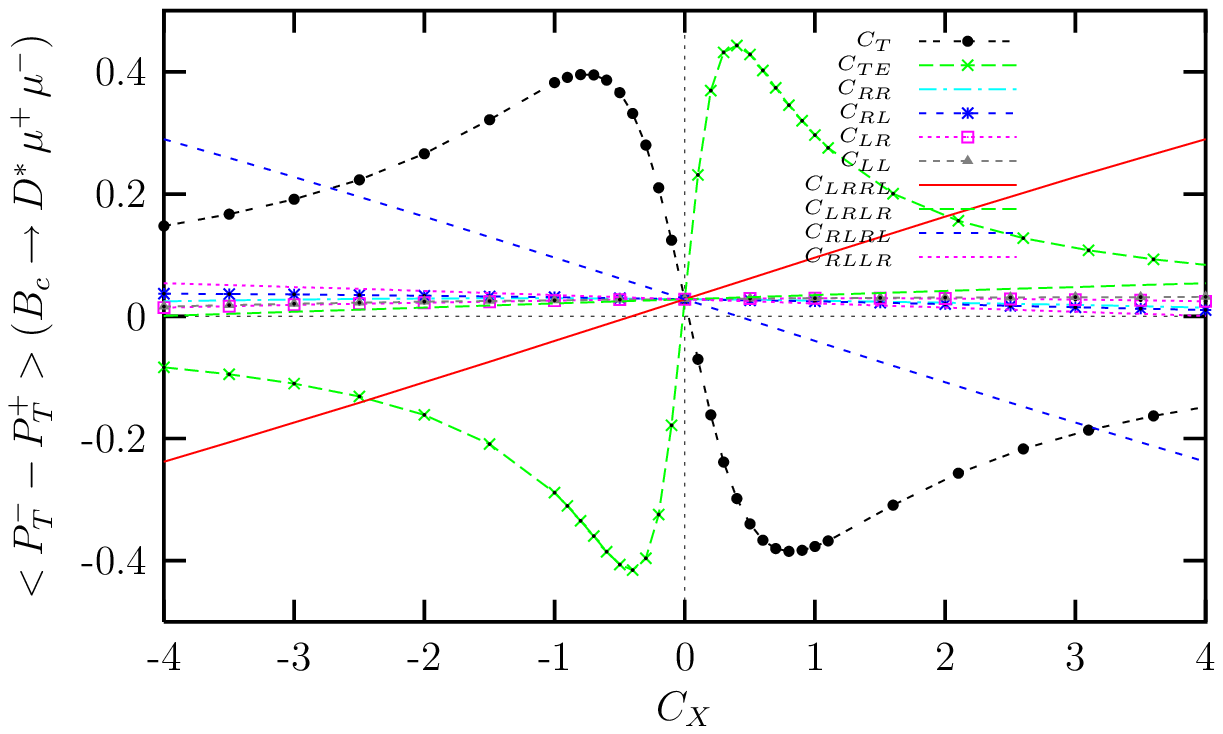}
\caption{The dependence of the combined averaged transverse lepton
polarization \, \, \, $<P^-_T-P^+_T>$ for the $B_c \rar
D_{s}^{\ast}\gamma \, \mu^+ \mu^-$  decay on the new Wilson
coefficients.\label{f8}}
\end{figure}
\clearpage
\begin{figure}
\centering
\includegraphics[width=5in]{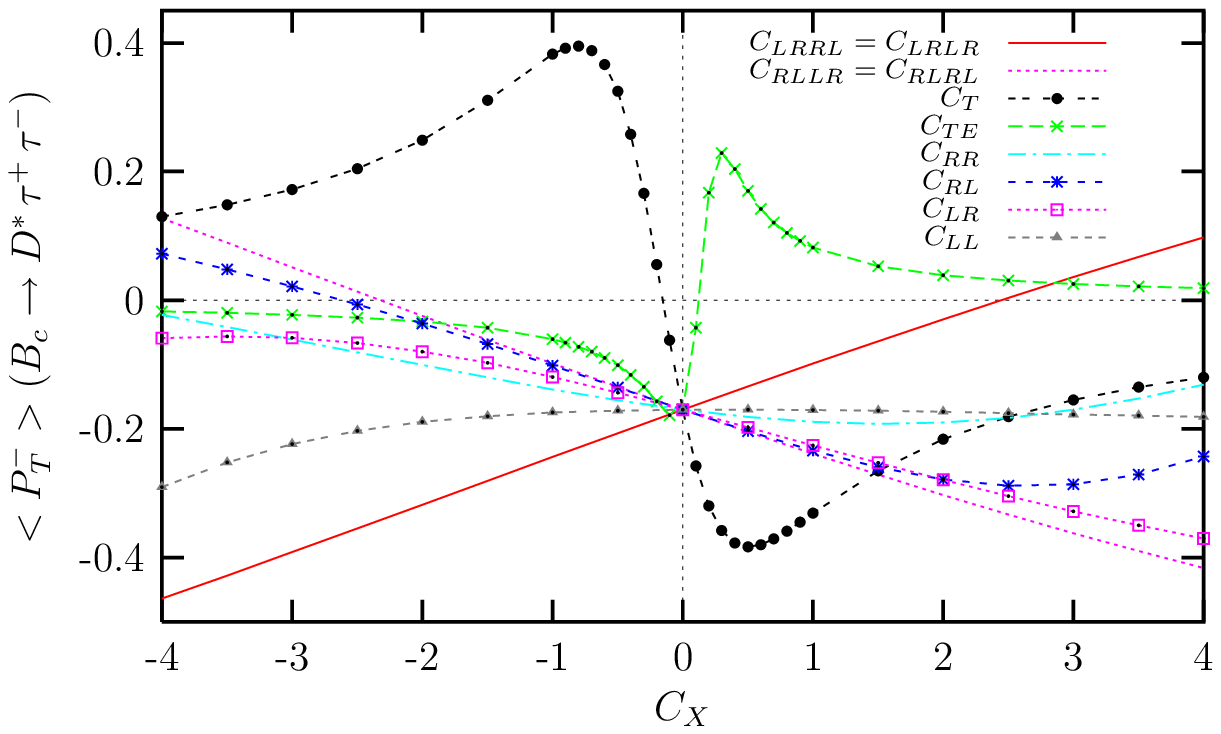}
\caption{The same as Fig. (\ref{f7}), but for the $B_c \rar
 D_{s}^{\ast}\, \tau^+ \tau^-$ decay. \label{f9}}
\end{figure}
\begin{figure}
\centering
\includegraphics[width=5in]{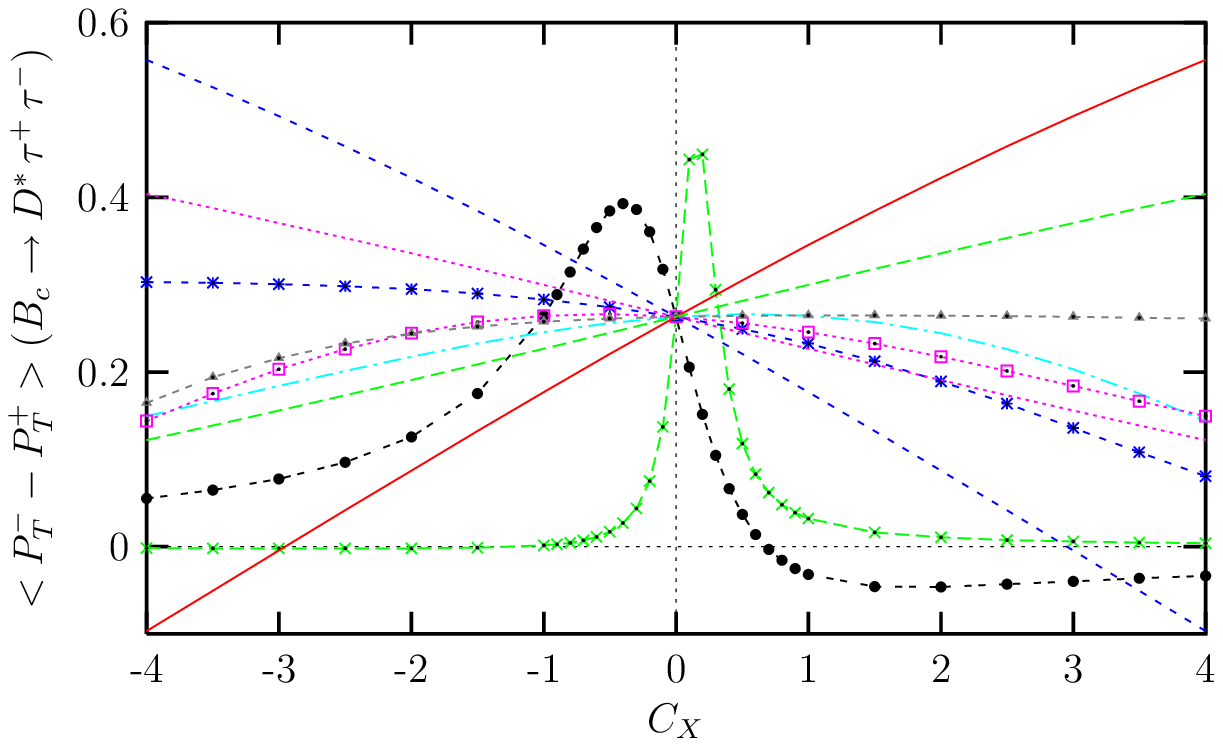}
\caption{The same as Fig. (\ref{f8}), but for the $B_c \rar
 D_{s}^{\ast}\, \tau^+ \tau^-$ decay.\label{f10}}
\end{figure}
\clearpage
\begin{figure}
\centering
\includegraphics[width=5in]{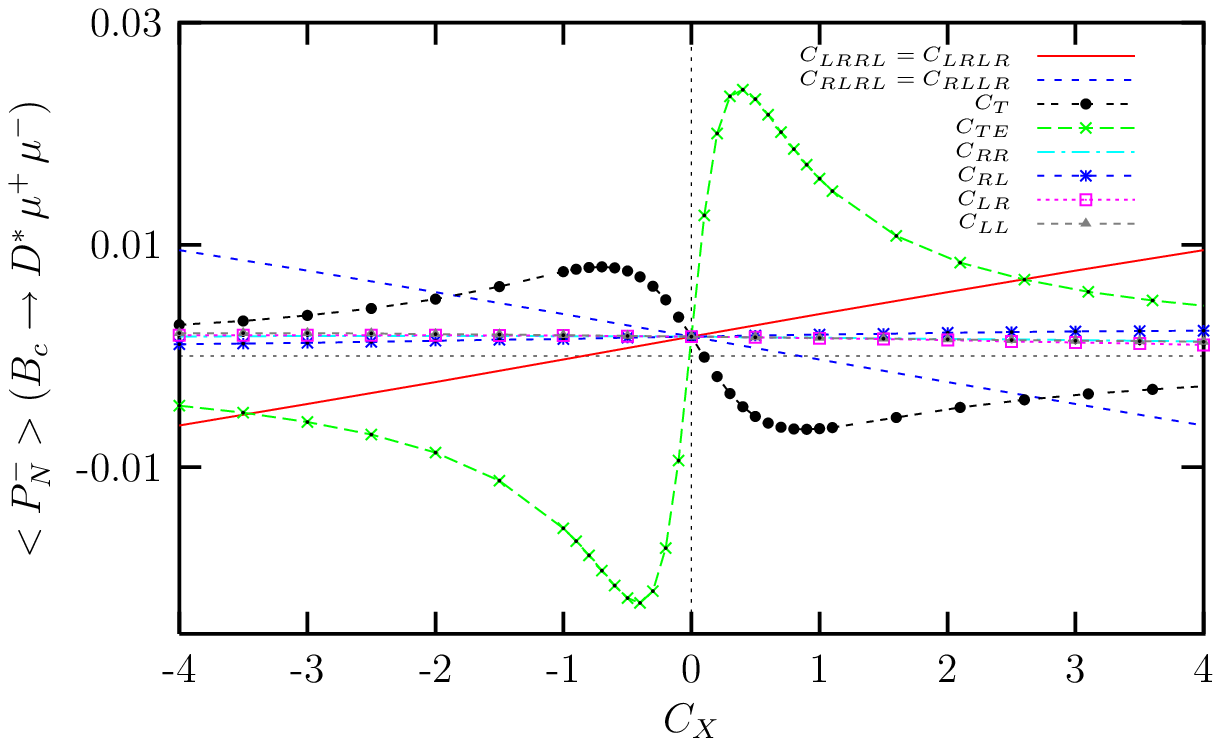}
\caption{The dependence of the averaged normal polarization
$<P^-_N>$ of $\ell^-$ for the $B_c \rar D_{s}^{\ast} \, \mu^+ \mu^-$
decay on the new Wilson coefficients.\label{f11}}
\end{figure}
\begin{figure}
\centering
\includegraphics[width=5in]{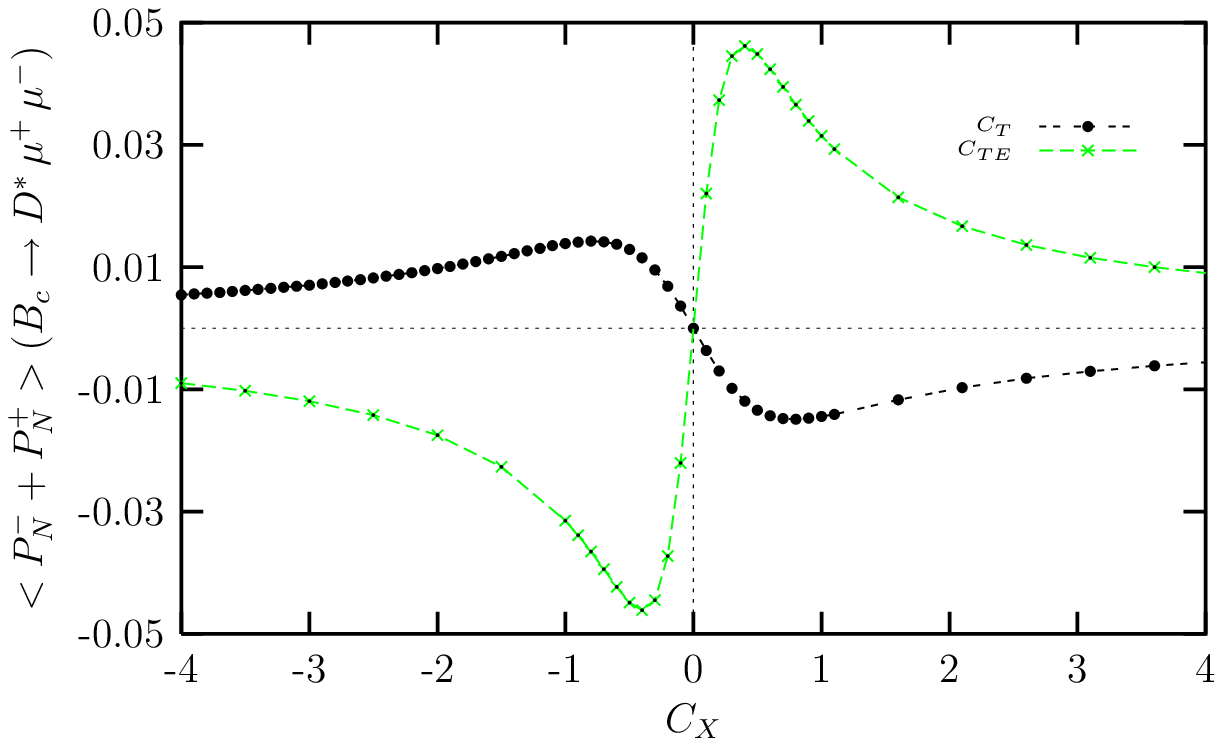}
\caption{The dependence of the combined averaged normal lepton
polarization \, \, \, \, $<P^-_N+P^+_N>$ for the $B_c \rar
D_{s}^{\ast} \, \mu^+ \mu^-$  decay on the new Wilson
coefficients.\label{f12}}
\end{figure}
\begin{figure}
\centering
\includegraphics[width=5in]{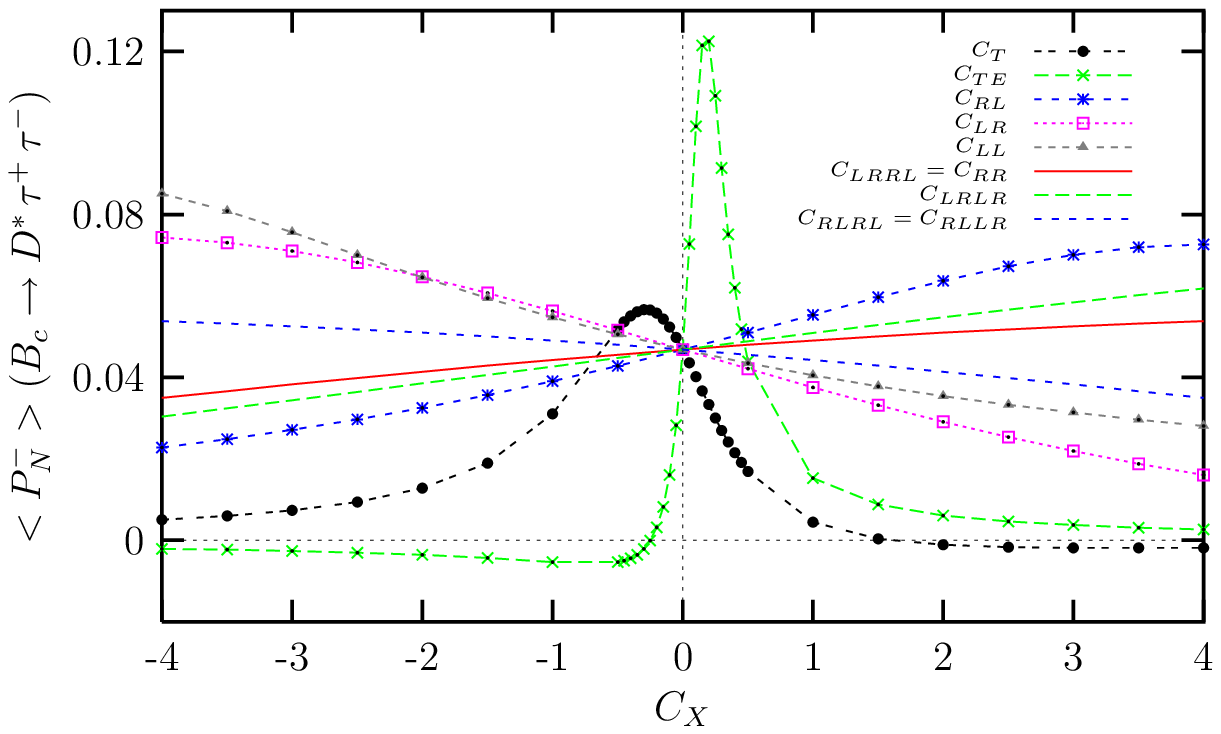}
\caption{The same as Fig.(\ref{f10}), but for the $B_c \rar
D_{s}^{\ast} \, \tau^+ \tau^-$  decay.\label{f13}}
\end{figure}
\begin{figure}
\centering
\includegraphics[width=5in]{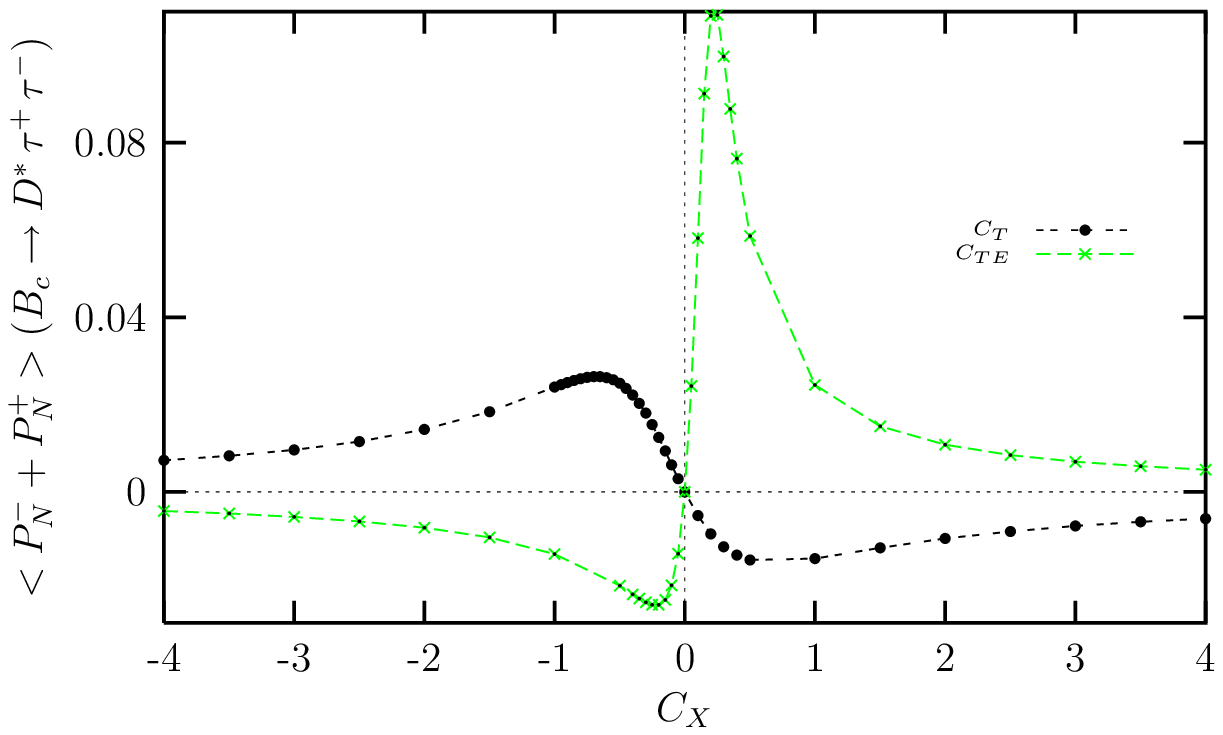}
\caption{The same as Fig.(\ref{f11}), but for the $B_c \rar
D_{s}^{\ast} \, \tau^+ \tau^-$  decay.\label{f14}}
\end{figure}

\end{document}